\newif\ifsubmit\submitfalse
\newif\ifrevtex\revtextrue
\ifrevtex
\ifsubmit                       
\documentstyle[epsf,journals_aps,subeqnarray,jprart,defs,prd,aps,amsfonts,preprint]{revtex}
\else                           
\documentstyle[epsf,RM98prd_japs,subeqnarray,RM98prd_jprart,RM98prd_defs,times,prd,aps,amsfonts,floats]{revtex}
\draft

\fi
\renewcommand{\gsim}{\gtrsim}   
\renewcommand{\lsim}{\lesssim}
\newcommand{\capsize}{\relax}
\else                           
\documentstyle[11pt,uslet,epsf,journals,subeqnarray,times,jprart,defs,amsfonts]{article}
\newcommand{\capsize}{\small}
\submitfalse                    
\fi

\begin{document}

\title{Photohadronic Neutrinos from Transients in Astrophysical Sources}

\ifrevtex
\author{J\"org P. Rachen and P. M\'esz\'aros}
\address{Department of Astronomy and Astrophysics,
	The Pennsylvania State University, University Park, PA 16802\\
	e-mail: jrachen@astro.psu.edu, pmeszaros@astro.psu.edu}
\ifsubmit\relax\else\date{Submitted to Phys.~Rev.~D: 17 February, 1998;
published 18 November, 1998}\fi
\else
\author{J\"org P. Rachen and P. M\'esz\'aros\\[1pc]
	Department of Astronomy and Astrophysics\\
	The Pennsylvania State University, University Park, PA 16802\\
	e-mail: jrachen@astro.psu.edu, pmeszaros@astro.psu.edu}
\date{Submitted to Phys Rev D, February 17, 1998}
\fi
	
\maketitle

\begin{abstract} 
We investigate the spectrum of photohadronically produced neutrinos at very
high energies (VHE, ${\gsim} 10^{14}\eV$) in astrophysical sources whose
physical properties are constrained by their variability, in particular jets
in Active Galactic Nuclei (blazars) and Gamma-Ray Bursts (GRBs).  We discuss
in detail the various competing cooling processes for energetic protons, as
well as the cooling of pions and muons in the hadronic cascade, which impose
limits on both the efficiency of neutrino production and the maximum neutrino
energy. If the proton acceleration process is of the Fermi type, we can
derive a model independent upper limit on the neutrino energy from the
observed properties of any cosmic transient, which depends only on the
assumed total energy of the transient. For standard energetic constraints, we
can rule out major contributions above $10^{19}\eV$ from current models of
both blazars and GRBs; and in most models much stronger limits apply in order
to produce measurable neutrino fluxes. For GRBs, we show that the cooling of
pions and muons in the hadronic cascade imposes the strongest limit on the
neutrino energy, leading to cutoff energies of the electron and muon neutrino
spectrum at the source differing by about one order of magnitude. We also
discuss the relation of maximum cosmic ray energies to maximum neutrino
energies and fluxes in GRBs, and find that the production of both the highest
energy cosmic rays and observable neutrino fluxes at the same site can only be
realized under extreme conditions; a test implication of this joint scenario
would be the existence of strong fluxes of GRB correlated muon neutrinos up
to ultra high energies, ${>}10^{17}\eV$.  Secondary particle cooling also
leads to slightly revised estimates for the neutrino fluxes from
(non-transient) AGN cores, which are commonly used in estimates for VHE
detector event rates. Since our approach is quite general we conclude that
the detection or non-detection of neutrinos above ${\sim}10^{19}\eV$
correlated with blazar flares or GRBs (\eg, with the Pierre Auger
Observatory), would provide strong evidence against or in favor of current
models for cosmic ray acceleration and neutrino production in these sources.
\end{abstract}

\ifrevtex
\fi

\sect{intro}{Introduction}

Neutrino astronomy may provide valuable clues for the understanding of the
properties of neutrinos and their interactions at energies in the range
$10^{14}{-}10^{19}$ eV, as well as providing qualitatively new information
about some of the most interesting cosmic objects. This energy range is also of 
great interest because it can probe the universe at significantly greater 
distances than is possible with known stellar sources (\eg, the Sun, or 
Supernovae such as 1987a, produce neutrinos in the ${\sim}\MeV$ range through 
nuclear interactions, which would be difficult to detect from more distant 
sources due to the overwhelming background of atmospheric neutrinos from cosmic 
ray air showers \cite{Tot92}).  For these reasons, many of the future neutrino
telescopes are designed for energies ${\gsim}\TeV$, where the atmospheric
background becomes negligible. Among the most promising and ubiquitous
astrophysical sources of neutrinos at these very high energies are Active
Galactic Nuclei and Gamma-Ray Bursts \cite{GHS95}, which have in common that
most of their energetic emission appears in short, distinct flares. The study
of the physical processes determining the energy spectrum of such neutrino
emitting transients is the subject of this paper. 

Above TeV energies and up to about $10^{17}$ eV neutrinos are detected
predominantly through the Cherenkov effect in large volumes of water or ice,
using the mass of the earth to capture neutrinos and looking for traces of
upward going muons from $\nu_\mu\to\mu$ conversions; above $10^{17}$ eV air
scintillation techniques and large air shower arrays become the most
efficient to detect neutrino induced, deeply penetrating horizontal air
showers, where electron neutrinos have the advantage to generate showers
which are easier to distinguish from the cosmic ray background (see
\RefApp{nu:event} for details and references). The major change of detection
techniques at about $10^{17}\eV$ motivates the distinction between very high
energy (VHE) neutrinos at $10^{14}{-}10^{17}\eV$ and ultra high energy (UHE)
neutrinos at ${\gsim}\,10^{17}\eV$. The most obvious source of UHE neutrinos
are interactions of ultra-high energy cosmic rays with the universal
microwave photon background \cite{Ste73/79}, which predict a diffuse
neutrino flux strong enough to be detected in air shower experiments
\cite{Zas97}.  More hypothetical is the prediction of UHE neutrinos from
processes associated with grand unification scale physics, \eg, the
annihilation topological defects \cite{SLSC97}. In the VHE range, the major
contribution is expected from Active Galactic Nuclei (AGN)
\cite{SDSS91,SB92,SP94,Man95app,SS96,HZ97}, which are known to emit very high
energy gamma-rays. A possible contribution of AGN to the UHE neutrino regime
is also discussed in connection with the expected event rates of horizontal
air showers \cite{PZ96}. Also Gamma-Ray Burst (GRB) sources have been
proposed as neutrino sources in the $\MeV{-}\GeV$ range \cite{PX94}, and at
very high energies \cite{WB97}.

Neutrino production in hadronic models of AGN and GRB is generally attributed
to the acceleration of protons in shocks or plasma turbulence, known as Fermi
acceleration. These energetic protons then interact with soft background
photons to produce pions (photohadronic pion production).  VHE/UHE neutrinos
originate in the decay of charged pions, boosted in energy by the original
proton Lorentz factor, and maybe by an additional Doppler factor due to the
bulk motion of the rest frame of the relativistic outflows or jets which
characterize these sources. The associated decay of neutral pions leads to
the production of gamma-rays, and it has been claimed that the observed
emission up to ${\gsim}\,10\TeV$ in Doppler boosted AGN jets
\cite{ICRC97-TeV} is due to this process rather than inverse Compton
scattering of photons by energetic electrons \cite{Man93/96}. A relevant
issue is that there are alternative models of AGN and GRB where the momentum
and energy flux of the relativistic jets or outflows are provided by $e^\pm$
or magnetic fields (\eg, \cite{SMB97,MRees97apjl} and references therein),
rather than protons, and these would be expected to have negligible neutrino
fluxes. The positive identification of VHE/UHE neutrinos from AGN and GRB
would be an indication for baryon loaded outflows. The energetic protons may
also contribute to the highest energy cosmic ray spectrum, which is observed
up to $3\mal10^{20}\eV$ \cite{Bie97jphG}, where AGN and GRBs are considered
among the most plausible sources. Here also, the observation or
non-observation of energetic neutrinos would be a crucial test for these
models.

The details of the photohadronic production of neutrinos via pion decay
depend strongly on the properties of the source, \ie, its size, lifetime,
magnetic field, etc. The magnitude of these quantities are estimated from the
observed variability, flux, and the measured or inferred distance of these
sources. GRB usually last $0.1{-}100$ seconds, but show intrinsic variability
down to milliseconds, while AGN emit most of their energetic radiation in
strong flares lasting several weeks, with intrinsic variability on time
scales of days down to less than one hour. The transience of energetic
emission could improve the association of detected neutrinos with their
putative sources, because one could use both arrival direction and arrival
time information, allowing statistically significant statements even for
total fluxes below the background level.

On the other hand, transience and variability sets constraints on the maximum
energy of the neutrino spectrum. In the literature so far, this has been
connected to the maximum proton energy using simple kinematical relations. As
we show in this paper, however, in these astrophysical scenarios the
secondary particles in the photohadronic cascade, \ie, pions and muons, have
to be considered separately, since cooling processes can have a significant
impact on their final distribution. Moreover, one needs to evaluate carefully
the competing proton energy loss processes that do not lead to neutrinos,
which can cause breaks in the neutrino spectrum that are not present in the
proton spectrum, and thus strongly limit the predicted fluxes at very high
energies.

Starting with a general treatment of photohadronic neutrino production in
variable sources, we derive a general upper limit for the maximum energy of
neutrinos produced by photohadronic interactions of Fermi accelerated protons
in cosmic transients, which only depends on the total energy of the transient
and observational parameters, like duration or (photon) luminosity. We then
apply our results to hadronic AGN and GRB models, and find that they impose
severe constraints on their possible contribution to the UHE neutrino
spectrum. Notational conventions used throughout the paper and frequently
used symbols are explained in \RefApp{notation}.

\sect{phothad}{Photohadronic neutrino production} 

\sub{pcool}{Proton cooling and neutrino production efficiency}

Photohadronic neutrino production is a result of the decay of charged pions
originating from interactions of high energy protons with ambient low energy
photons. It is accompanied by the production of gamma-rays from neutral pion
decay; the details of the process are described in \RefApp{phothad}, for a
target photon spectrum following a power law with index $a$ above a break
energy $\epsb$. For the proton gas, pion production acts as a cooling
process, and is in competition with other cooling processes like
Bethe-Heitler pair production, synchrotron radiation, cosmic ray emission,
and adiabatic losses due to the expansion of the emission region.

While neutrinos are exclusively produced by charged pion decay, gamma-rays
are produced by a variety of processes: besides neutral pion decay, $\piO \to
\gamma\gamma$, the major hadronically induced channels are synchrotron
radiation from (a) the UHE protons themselves, (b) the electrons, muons and
charged pions in the photohadronic decay chain (see \RefApp{phothad}), and
(c) the Bethe-Heitler pairs produced in $p\gamma\to p\ep\ee$ interactions
\cite{BG70}. If synchrotron cooling of secondary particles is negligible,
about $25\%$ of the energy in charged pions is converted into gamma
radiation by synchrotron cooling of the electron produced in muon decay
\cite{Man93/96,MKB91}. Normally, the first generation gamma-rays cannot leave
the emission region, but rather induce an electromagnetic cascade through
pair production with low energy background photons, and subsequent
synchrotron radiation of electrons and positrons. They cascade down in
energy, until they eventually escape below some critical energy where the
emission region becomes optically thin \cite{MKB91}. Hadronically induced
gamma-rays are usually in competition with synchrotron and inverse Compton
photons radiated by primary energetic electrons.

Cosmic rays can be ejected in essentially two ways: (a) if the emission
region has a sharp boundary beyond which the magnetic field drops rapidly,
protons scattered across the boundary would be ejected, and (b) secondary
neutrons produced in $p\gamma$ interactions can escape if (b1) their decay
length in the comoving frame is larger than the size of the emission region,
$c\tau_n \gg R$ with $\tau_n = \gamma_p\taunRF$, and (b2) their probability
of re-conversion to a proton by, \eg, a reaction $n\gamma\to p\pim$ is small,
expressible by $c\tnp \gg R$. Process (a) depends on the detailed structure
of the emission region and is usually in competition with adiabatic cooling,
which affects charged particles due to the adiabatic invariance of the
quantity $B\rgyr^2$ during the Larmor motion of the particle ($\rgyr = E/eB$
is the Larmor radius) in a magnetic field decreasing with expansion
\cite{Jackson}. In an isotropically expanding emitter with conserved total
magnetic energy this means $B\propto R^{-2}$, and thus $E\propto R^{-1}$, but
other dependences may apply (see \RefSec{theory:Epmax:larmor}).  Process (b)
is tightly connected to neutrino production, because the dominant channel
producing charged pions, $p\gamma \to n\pip$, is also the dominant channel
for neutron production. The time scale for proton-neutron conversion is $\tpn
\approx \tnp \approx 2\tpi$, which is larger than the time scale for charged
pion production, $\tpipm \approx \frac32 \tpi$ (\RefApp{phothad}), because
there the process $p\gamma\to p\pip\pim$ contributes considerably.

The efficiency of neutrino production depends on (i) which fraction of their
energy protons convert into charged pions, and (ii) the fraction of energy
pions and muons retain until they decay. Condition (ii) can be quantified by
introducing the efficiency of energy conversion from the originally produced
pion into neutrinos, $\Xnumu = \frac12 (\gmudec/\gpiprod)$ and $\Xnupi =
\frac14 (\gpidec/\gpiprod)$, for muon and pion decay, respectively, where
$\gpiprod \approx \gamma_p$ is the Lorentz factor of the pion at production,
and $\gmudec\le\gpidec\le\gpiprod$ are the Lorentz factors of the muon and
the pion, respectively, at their decay (see \RefApp{phothad} and
\RefSec{theory:pimu}). Similarly we can quantify (i) by introducing the
charged pion production efficiency, $\Xpipm = \tpcool/\tpipm$, where
$\tpcool$ is the total cooling time of the proton. This leads to the total
neutrino production efficiency
\eqn{Xnu}
\Xnu(\gamma_p) = \Xpipm(\Xnupi + \Xnumu) \approx
\left(\frac13\,\frac{\gmudec}{\gamma_p} +
\frac16\,\frac{\gpidec}{\gamma_p}\right)\frac{\tpcool}{\tpi} \quad.
\text
The total proton cooling time is determined by the inverse sum, $\tpcool^{-1}
= \sum_i t_{p,i}^{-1}$, extending over all participating cooling
processes. To classify the cooling processes by their dependence on
$\gamma_p$, we introduce total cooling times for photohadronic interactions,
$\tpg$, synchrotron radiation, $\tsynp$, and external cooling processes,
$\tec$.  Under external cooling processes we subsume adiabatic cooling, which
has a time scale $\tad$ independent of the proton energy, and direct ejection
of protons from the emission region. The latter may be dependent on the
proton energy if diffusive losses are relevant; in the simplest case,
however, we can assume that protons are confined over the time scale set by
adiabatic expansion, \ie, $\tescp \gg \tad$, which means that
\eqn{tad}
\tec \approx \tad = \const(\gamma_p)\quad.
\text
The synchrotron loss time can be written as
\eqn{tsyn}
\tsynp = \tsynb \left(\frac{\gamma_p}{\gb}\right)^{-1} 
\with \tsynb = \frac{9c}{4\rp\,\omBp^2\gb}\;,
\text
where $\rp = e^2/m_p c^2 \approx 1.5\mal 10^{-16}\cm$ is the classical proton
radius, and $\omBp = e B/m_p c$ is the cyclotron frequency of the proton. The
characteristic proton Lorentz factor used here for normalization,
\eqn{gb}
\gb \equiv \eth/2\epsb
\text
expresses the limit above which all photons in the power law part of the
spectrum are boosted above the reaction threshold for pion production,
assuming that the photon number spectrum can be described as a power law,
$d\Nph \propto \eps^{-a}\,d\eps$, with an index $a>1$ above some break energy
$\epsb$ (see \RefApp{phothad} for details). The cooling time for pion
production can then be written in a similar way as
\eqns{tpi}
\tpi &=& \mlbox{10}{\tpib (\gamma_p/\gb)^{1-a}} \for\gamma_p < \gb\\
\tpi &\approx& \mlbox{10}{\tpib (\Ngb/\Ngtot)}  \for\gamma_p \gg \gb\quad,
\text
where $\Ngtot$ is the total photon density, $\Ngb$ is the density of photons
with $\eps>\epsb$.  $\tpib$ is the pion production cooling time for protons
with $\gamma_p = \gb$, and can be expressed as $\tpib = [c \Ngb
\Etapi]^{-1}$, where $\Etapi$ is the inelasticity weighted effective cross
section for pion production, as defined in \RefApp{phothad}.

The time scales for other photohadronic cooling processes, including neutron
ejection, can all be expressed in $\tpi$. The cooling time for the
Bethe-Heitler process can be evaluated similarly to
\eqs{tpi} and \RefEq{APPENDIX:phothad:Etapi}; the inelasticity weighted cross
section for the Bethe-Heitler process is $\EtaBH \approx \Etapi/125$, for a
break Lorentz factor $\gbBH \approx \gb/140$, leading to $\fBH \equiv
\tpi/\tBHp \approx \exp(5a-10)$ for $\gamma_p \lsim \gbBH$, and $\fBH\ll 1$
for $\gamma_p \gg \gbBH$. To quantify the time scale for energy loss due to
free neutron escape, $\tescn$, we introduce the probabilities for
neutron-to-proton reconversion within the length scale of the emission region
$R$, due to beta decay, $\Pbeta = \exp(-c\tau_n/R)$, and due to pion
production, $\Png \approx \exp(-2 c \tpi/ R)$, using $\tnp\approx 2\tpi$.
Since the typical energy ratio of the neutron to the pion in a $p\gamma\to
n\pip$ reaction is ${\approx}4$, we finally get $\tescn \approx \frac12
\Pescn^{-1} \tpi$, where $\Pescn = (1-\Png)(1-\Pbeta)$ is the probability of
the neutron to escape. With $\tpg^{-1} \equiv \tpi^{-1} + \tescn^{-1} +
\tBHp^{-1}$, and using the fact that beta decay and photohadronic
interactions for neutrons are likely in different Lorentz factor regimes,
\viz, $\Pbeta\Png\ll 1$, we can write
\eqn{fpg}
\fpg \equiv \frac{\tpi}{\tpg}  \approx 3 + \exp(5 a - 10) 
	- 2\exp\left(-\frac{c\taunRF\gamma_p}R\right) 
	- 2\exp\left(-\frac{2 c\tpi}{R}\right)
\text
for $\gamma_p\lsim 0.01 \gb$. For $\gamma_p \gsim \gb$, the term expressing
the Bethe-Heitler efficiency, $\exp(5a-10)$, is absent. Assuming $\tec\approx
\tad$ we can write
\eqn{tcool/tpi}
\frac{\tpi}{\tpcool} = \Big(\tpg^{-1} + \tsynp^{-1} + \tec^{-1}\Big)\tpi
	\approx \max(\fad,\fpg,\fsyn) \equiv \fmax\;,
\text
where $\fsyn \equiv \tpi/\tsynp$ and $\fad \equiv \tpi/\tad$ are defined
analogously to $\fpg$, \ie, expressing the energy dissipated in the various
cooling channels in units of the energy lost in pion production. The
approximation by the maximum-function is best if one cooling process clearly
dominates, and is useful for the following, qualitative discussion of the
spectral distribution of emitted neutrinos.

\sub{specshape}{Shape of the time integrated neutrino spectrum}

Because of the low detection efficiency of neutrinos at earth, it is
impossible with present techniques to observe short scale time variability of
cosmic neutrino spectra. Therefore, for an outburst active over a limited
time, it is more meaningful to calculate the time integrated neutrino count
spectrum, rather than the spectral count rate at a fixed time. This also
simplifies the theoretical treatment, because we do not need to perform a
self consistent calculation of the accumulated proton spectrum at a specific
time --- the energy input at any specific energy is simply given by the time
integrated proton {\em injection} spectrum.

Clearly, the proton spectrum injected by the acceleration process is not
directly observable. We will follow here the scenario assumed in most models,
that the average injection rate for energetic protons follows a power
law in energy, $\exval{d\Npdot} = I_p \gamma_p^{-s} d\gamma_p$, extending
from some minimum Lorentz factor $\gpmin$ to a maximum Lorentz
factor $\gpmax\gg\gpmin$. We assume that the injection is active over a
time $\Tinj$, and that the injection spectrum does not change in time. Then
the total, time integrated energy density of injected protons is given by
\eqn{Up}
\Upint = \Tinj\int_{\gpmin}^{\gpmax}\! 
	m_p c^2 \gamma_p \exval{d\Npdot}_{\gamma_p}  = \bp m_p c^2 \Tinj I_p
	\gpmax^{2-s},
\text
where $\bp$ is the bolometric correction factor of the proton spectrum
relative to its energy content at the highest particle energies, given as
$\bp = |(\gpmax/\gpmin)^{s-2}-1|/|s-2|$ for $s\neq 2$ and
$\bp=\ln(\gpmax/\check\gamma_p)$ for $s=2$.

Protons injected at a specific Lorentz factor $\gamma_p$ produce charged
pions with $\gpiprod\approx \gamma_p$ at a rate $\tpipm^{-1}$, over a time
$\tpcool$. Thus, the total number density of charged pions produced in the
emission process is $d\Npi = \frac 23 \Tinj \exval{d\Npdot} \tpcool
\tpi^{-1}$.  Each charged pion produces two muon neutrinos
($\nu_\mu\bar\nu_\mu$) and one electron neutrino ($\nu_e$ or $\bar\nu_e$),
each with an energy $\Enu \approx \frac14 m_\pi c^2 \gamma_p$ if we assume
that $\gmudec \approx \gpidec \approx \gamma_p$.  Then, the total time
integrated neutrino power at the energy $\Enu$ emitted by the source in its
rest frame is
\eqn{specnu}
\LTnu(\Enu) \equiv \Enu\;\frac{d\Nnu}{d\ln\Enu} =
\frac{m_\pi}{m_p}\,\frac{V \Upint}{\bp\gpmax^{2-s}}\;
	\Phi\!\left(\frac{4\Enu}{m_\pi c^2}\right)\quad, 
\text
where $V$ is the volume of the emission region. The spectral shape is
expressed by the function $\Phi(\gamma_p) = \gamma_p^{2-s}\Xnu(\gamma_p)$,
which can be written as $\Phi(\gamma_p) = \gamma_p^{2-s}\fmax^{-1}(\gamma_p)
\propto \gamma_p^{q}$ in the case of one dominating cooling process; the
power law index $q$ is, depending on the dominant cooling process, given by
\keepeqno\eqns{specnu}\subeq{q}
\begin{array}{l@{\;\;:\quad}l@{\qquad}l}
\Phi(\gamma_p)\propto\gamma_p^q & \quad\gamma_p \lsim \gb & \gamma_p \gg \gb
\\\hline 
\fmax = \fad\quad & \quad q=a-s+1 & q=2-s \vphantom{\bigg|} \\ 
\fmax = \fpg  & \quad q=2-s & q=2-s \\ 
\fmax = \fsyn & \quad q=a-s & q=1-s \vphantom{\bigg|} \\
\end{array}\qquad,
\text
where $a$ is the target photon spectral index.  For $a>2$, an additional
spectral modification will occur due to the drop of the Bethe-Heitler
efficiency between ${\sim}\,0.01\gb$ and ${\sim}\,\gb$, if photohadronic
cooling is dominant in this region ($\fmax=\fpg$); in this case, one
would expect a rapid rise in the neutrino flux in the regime $\Enu \lsim
[30\MeV]\;\gb$. 

The energy dependence of neutrino event rate as observed in a given detector
follows closely the neutrino power spectrum, as shown in \RefApp{nu:event},
if we properly account for energy shifts due to Doppler boosting or source
redshift, \ie, we expect an increase of events with energy for $q> 0$, a
decrease for $q< 0$, and $q\approx 0$ would indicate an event rate almost
independent of energy. The dominant proton cooling process in the source, and
thus the value of $q$ may change with energy. Obviously, such a transition of
cooling processes at some energy, leading to a spectral break, can only occur
if the process taking over has a cooling time decreasing faster with energy.
Therefore, the spectrum steepens at each break, and for $\gpmax<\gb$ the only
possible sequence of break energies is $\Enubpg\le\Enubpsyn\le\Epmax$ for
$a<2$, and $\Enubpsyn\le\Enubpg\le\Epmax$ for $a>2$, while only one break can
exist for $a=2$ or above $\gb$ [\cf\ \eq{specnu:q}]. However, depending on the
source properties it may be that some cooling processes are never dominant,
so not all possible break energies may appear in the spectrum [\cf\
\RefSec{astro:AGN} and \Fig{astro:AGN:parspace} for an example].  If the
proton spectral index is close to the canonical value for shock acceleration,
$s\approx 2$, we would generally expect an increasing event rate at low
energies in the regime of adiabatic cooling dominance, and a flat behavior
($q\approx 0$) if photohadronic cooling becomes dominant, which can be most
easily understood as a saturation of the efficiency, \eq{Xnu}. The efficiency
can decrease again if proton synchrotron cooling becomes dominant (for
$a<2$), unless the proton spectrum cuts off first. The possible cooling of
secondary particles in the hadronic cascade may lead to additional breaks, as
discussed in \RefSec{theory:pimu:specmod}, and illustrated in an example in
\RefSec{astro:GRB} and \Fig{astro:GRB:parspace}.

\sect{theory}{Neutrino emission from cosmic transients: general theory}

Although neutrino bursts themselves may only be observable in their time
integrated appearance, the accompanying burst of photons is observable in
much greater detail, and allows to constrain the physical parameters
determining neutrino production. We have to distinguish between low energy
photons, which are in general explained by synchrotron radiation of electrons
co-accelerated with the protons, and high energy gamma radiation, which may
be dominated by hadronically induced cascades, as discussed above, but could
also originate dominantly from inverse Compton photons produced by the
electrons. The terms ``low energy'' and ``high energy'' are here used only in
a relative meaning --- the absolute energy range for electron synchrotron
radiation on the one hand, and electron Compton radiation or hadronically
induced photons on the other, depends strongly on the physical conditions at
the source.  The following discussion is more focussed on the low energy
photon component, which is relevant as the target population for
photohadronic neutrino production. However, if the high energy photons are of
hadronic origin, their variability can also give valuable clues on proton
cooling times.

Hereafter we distinguish between physical quantities defined in the comoving
frame of the source and observed quantities (see \RefApp{notation} for
notational conventions), where we account for a possible boosting of the
radiation emitted from the source with a Doppler factor $\cD =
[\Gamma(1-\bflow\cos\phiview)]^{-1}$ for a relativistic flow with Lorentz
factor $\Gamma$ and a velocity $\bflow c$ under an angle $\phiview$ to the
direction of the observer. We do not take into account cosmological redshift
effects, and just note that they might be considered by replacing $\cD =
\cD'/(1+z)$, if $\cD'$ is the Doppler factor of the emission region in the
cosmologically comoving frame at the source.  The task of this section is to
constrain the physical properties of the emission region by observable
quantities, in order to discuss the various processes limiting the neutrino
energy. 

\sub{times}{Variability time scales and the size of the
emission region}

\subsub{causality}{The causality limit}

If a flare occurring in a relativistic outflow, boosted with a Doppler factor
$\cD$, is observed to have a duration $\cT$, the time scale of the burst in
the comoving frame of the fluid is $T = \cT\cD$. This flare time scale covers
(a) the time scale for the injection of energetic particles, $\Tinj$, (b) the
time scale over which the particles convert their energy in radiation,
$\Trad$, and (c) the crossing time the photons need to leave the emission
region in the direction of the observer, $\Tcr$. The partial times normally
do not simply add up, but by order of magnitude the estimate 
\eqn{Ttot}
T \sim \max(\Tinj,\Trad,\Tcr)
\text
applies. The crossing time is naturally connected to the (comoving) linear
size of the emission region; if the emission region is not spherically
symmetric, we can only limit the comoving size along the line of sight,
$\Rpar$, by the observed duration as
\eqn{Rlim}
\Rpar = c \Tcr = c \rhoT \cT \cD\quad,
\text
where the factor $\rhoT\le 1$ considers the effect of a delayed emission due
to finite injection or radiation time scales; $\rhoT=1$ means that the
emission is homogeneous and instantaneous within the size $R$ --- the
observed duration is then simply the time between the first and the last
photon reaching us. Since the condition $T\ge \Tcr$ is equivalent to the
requirement that the emission throughout the emission region is due to one,
causally connected process, \eq{Rlim} may be called the {\em causality limit}
for the size of the transient source.  The projected (or lateral) comoving
size, $\Rperp$, is not constrained by the variability time scale, but plays a
role for the determination of the internal radiation density of the emission
region from its observed, isotropized luminosity at a specific photon energy,
$\cL(\veps) = 4\pi \dL^2 \veps^2 (d\Nph/d\veps)_{\rm obs} = \cD^4 L(\eps)$
with $\veps = \cD \eps$, where $\dL$ is the luminosity distance of the source
(note that the redshift is absorbed in the Doppler factor, as explained
above). To account for this, we introduce a geometrical eccentricity
parameter, $\xL$, by writing the luminosity as
\eqn{lum}
L(\eps) = 4\pi\Rpar^2 c \xL \,  \eps^2 (d\Nph/d\eps) \quad,
\text 
where $\eps^2 (d\Nph/d\eps)$ is the specific energy density of photons with
energy $\eps$ in the rest frame of the flow. For a spherically symmetric
emission region $\xL = 1$, while a disklike emission region
($\Rpar\ll\Rperp$) would be described by $\xL {\sim} (\Rperp/\Rpar)^2 {\gg} 1$.

The radiation time scale is obviously equivalent to the cooling time scale of
the radiating particles; for radiation processes involving electrons, it is
usually very short, $\Trad\ll\Tcr$, justifying its neglect. If we consider
radiation produced in photohadronic interactions, we can write
$\Trad\approx\tpcool$, since the time scale over which the electromagnetic
cascade evolves can be considered as short compared to $\tpcool$. $\Tinj$ is
the time over which an acceleration process is active, \eg, the lifetime of a
shock. Obviously, this sets a limit on the acceleration time of the
particles, $\tacc < \Tinj < \cT\cD$. Also here, this is barely relevant for
electrons, but sets an important limit for protons.  

\subsub{intshock}{The internal shock scenario}

As an example how observed time scales in transient emission phenomena may be
connected to the size of the emission region, we discuss the scenario of
energy conversion in relativistic flows by internal shocks. This scenario was
suggested originally by Rees \cite{Rees78mnras} for AGN jets, and was later
also applied to Gamma-Ray Bursts \cite{ReesM94}. 

We consider two plasma blobs of similar mass and density emitted within an
unsteady flow at times $t_1$ and $t_2$, $\Delta t = t_2-t_1 > 0 $, with
respective Lorentz factors $\Gamma_1$ and $\Gamma_2$, $\Gamma_2/\Gamma_1
\gsim 1$, \ie, the second blob has a larger velocity and thus catches up
with the first after some time $\sim \Gamma_1\Gamma_2\,\Delta t$. Assuming
that their relative velocities are supersonic, two strong shock waves moving
in opposite directions form when the blobs merge; they are called {\em
forward shock} and {\em reverse shock}, respective to their direction of
motion relative to the flow. In their center of mass frame (CMF), which has a
Lorentz factor $\Gamma \approx \sqrt{\Gamma_1\Gamma_2}$ in the observers
frame, the shocked material in region between the two shocks is at rest, and
is the source of the radiation. The shocks move each with a velocity $\bsh
c$, $\bsh\approx\sqrt{1-\Gamma_1/\Gamma_2}$, corresponding to a internal
shock Lorentz factor $\Gsh \approx \sqrt{\Gamma_2/\Gamma_1}$. The linear size
of the emitter in the direction of the flow, after the merging is complete,
is $\Rpar = 2 \Rpar'/\chi_\rho$, if $\Rpar'$ is the length of the blobs in
this direction in their respective rest frames, and $\chi_\rho =
4\Delta\Gamma + 3$ is the compression factor \cite{BMcK76}. Therefore,
$\Tinj\approx \Rpar'/c\chi_\rho\bsh = \Tcr/2\bsh$, which is the time each
shock needs to cross half this distance. For transrelativistic internal
shocks, $\bsh\approx\frac12$, the crossing time is therefore a good measure
for the injection time scale. The total efficiency for the dissipation of
energy by the shocks is given by $\Xish = 1-2\Gsh/(1+\Gsh^2)$, and is about
$20\%$ for $\Gsh\approx 2$.

We now assume that the radiation time scale is much shorter than the dynamic
time scales involved in the shock merging. Then, the emission follows closely
the motion of the shocks; if the observer is placed at an angle $\phiview\ll
1$ to the flow direction, the emission of the forward shock appears as a peak
of duration $\Tf = \Tinj \cD^{-1} (1-\bsh)$ in the observer frame, while the
emission of the reverse shock causes a peak with a duration $\Tr = \Tinj
\cD^{-1} (1+\bsh)$ and comparable total energy. For $\bsh\approx\frac12$, the
two peaks can thus have different lengths (but of the same order of
magnitude), and the total, superposed peak might appear asymmetric, with a
rise time $\Trise \approx \Tf/2\cD$, and a decay time $\Tdecay \approx
(\Tr-\frac12\Tf)/\cD$.  The crossing time is then correctly estimated by
$\Tcr = \cT \cD$ (or $\rhoT =1$), if we define
\eqn{Tcr}
\cT \equiv \Tr - \Tf \approx \Tdecay-\Trise\quad.
\text
We stress that this result is {\em independent of $\bsh$}. \eq{Tcr} makes use
of two assumptions: (a) that the two plasma blobs have comparable densities,
and (b) that $\Trad\ll\Tcr$. Condition (b) is mostly fulfilled, if we
consider synchrotron of inverse radiation from energetic electrons. Condition
(a) can be assumed to nearly fulfilled in internal shocks --- which is the
most important difference to {\em external} shocks, where the densities are
usually very different. If (a) is not fulfilled, forward and reverse shock
have different velocities in the CMF, and also different efficiencies in
energy conversion \cite{BMcK76,SP95}: hence, the forward and backward peak
have very different strengths, and the correlation of $\Tf$ and $\Tr$ with
$\Trise$ and $\Tdecay$ is less straightforward. If (b) is not fulfilled, \ie,
if $\Trad>\Tcr$, we expect $\Trad \sim \Tdecay\cD$, and $\Tcr \sim
\Trise\cD$. A similar situation arises for fast cooling, but the presence of
a secondary acceleration process which is not associated to the shock waves
(\eg, second order Fermi acceleration, see \RefApp{Fermi}), which can keep up
a population of energetic particles homogeneously over the region of the
shocked gas, and thus extend the emission as in the case of slow
cooling. Despite these ambiguities, we may assume that for flares with
considerable asymmetry, $\Tdecay-\Trise > \Trise$, \eq{Tcr} gives a reliable
upper limit on the crossing time. We also note that the geometrical
eccentricity parameter, $\xL$, as introduced in \eq{lum}, satisfies the
relation $\xL = \chi_\rho^2 \xL'$, if $\xL'$ is the eccentricity of the blobs
before they merge. Since $\chi_\rho^2\gsim 10$ for transrelativistic shocks,
this may give rise to assume rather disklike geometries in the internal shock
scenario; this conclusion, however, may not be over-interpreted because we
can obviously not rule out that the blobs have been originally elongated in
the flow direction, \ie, $\xL'<1$. The internal shock mechanism can readily
be applied to spherical (or quasi-spherical) outbursts, where the up-catching
``blobs'' have to be replaced by shells emitted with different velocities at
different times \cite{ReesM94}. We discuss this scenario, and its implication
for the geometrical factors $\rhoT$ and $\xL$, in more detail in
\RefSec{astro:GRB:fireball}.

\sub{Epmax}{Maximum energy of accelerated protons}

The predominance of power laws in nonthermal emission spectra suggests that
the radiating particles gain their energy by a stochastic process. Based on
an original idea of Fermi \cite{Fer49/54}, the most commonly discussed
stochastic acceleration processes fall into two parts: (a) {\em first order
Fermi-acceleration} by diffusive scattering of particles across strong shock
waves, also called {\em shock-acceleration} \cite{Dru83} and (b) {\em second
order Fermi-acceleration}, where the particles gain energy from the
scattering at plasma waves \cite{MelroseII,Kul79}. Since plasma waves are
responsible for the scattering, and thus for the diffusive motion of of
particles in shock acceleration as well, it is most likely that both
processes combine if strong shock waves are present \cite{Kru92}. Fermi
acceleration is assumed to be the dominant energy dissipation mechanism in
AGN cores and jets, and in Gamma-Ray Bursts. Since Fermi acceleration works
independent of particle mass and charge, any protons or ions present at the
shock should be accelerated as well as electrons. It has been claimed for
various classes of objects that this could be the generating process of the
observed cosmic ray spectrum, up to the highest energies of order
$10^{20}\eV$ \cite{Bie97jphG}. Here we discuss the maximum energies of
protons Fermi-accelerated in emission regions constrained by variability. The
time scale for Fermi acceleration is expressed as a multiple of the Larmor
time of the particle, $\tacc \equiv \thF \tgyr = 2\pi\thF\rgyr/c$, where we
assume $\thF$ as constant for simplicity; as discussed in \RefApp{Fermi},
this is only true for special assumptions on the diffusion coefficient (\eg,
Bohm diffusion). Concerning its magnitude, $\thF\gg 1$ applies in the most
cases, but $\thF\sim 1$ is probably possible for acceleration at relativistic
shocks (\RefApp{Fermi}).

\subsub{larmor}{Larmor radius and adiabatic limits}

To be accelerated up to an energy $\Epmax$ by a Fermi mechanism, we have to
require that the protons can be magnetically confined in the emission region,
\viz, $\Epmax \le e B \Rmin$, $\Rmin = \min(\Rpar,\Rperp)$. Using the
relations $\Rpar < c\cT\cD$, $E_\nu \le \frac14 m_\pi c^2 \gamma_p$, and
$\Enuobs = E_\nu \cD$, we obtain the limit
\eqn{Enumax:Larmor}
\Enumaxobs \le \Enumaxobsgyr = \frac{e m_\pi c}{4 m_p}\;B\cT\cD^2\rhoT\cgyr
\quad.
\text
The factor $\cgyr<1$ considers that usually only a limited fraction of the
effective size $\Rmin$ of the emission region can practically be used for
particle gyration; we will assume $\cgyr\lsim\frac13$ in the following.

Similar limits can be derived from acceleration time constraints arising from
the dynamic time scales involved in the acceleration process. The first
condition of this kind is $\tacc<\Tinj$, which can be specified in the
internal shock scenario as $\tgyr < \Tcr/2\thF\bsh$, and is obviously
equivalent to \eq{Enumax:Larmor} with $\cgyr = (2\thF\bsh)^{-1}$. The second
condition is the limitation of the acceleration time by adiabatic cooling in
a decreasing magnetic field, $\tacc<\tad \equiv 2 |B/\dot B|$. In an
expanding emission region, we usually find $B\propto R^{-\alpha}$, where
$\alpha > 0$ and $R$ is some characteristic size of the emission region. In
the general case, in particular for non-isotropic expansion, $\alpha$ may
depend on the choice of $R$; in an isotropically expanding emitter with
conserved magnetic energy we have $\alpha = 2$, which we may use as a
canonical assumption hereafter.  Defining the velocity of expansion as $\bad
\equiv \dot R/c$, this results again in \eq{Enumax:Larmor}, with $\cgyr =
(\pi\alpha\thF\bad)^{-1}$.  Adiabatic cooling is most relevant in a freely
expanding relativistic fluid, $\bad\approx 1$, or for rapidly decaying
magnetic fields, $\alpha\gg 1$ (however, we have to assume that the field
decay is adiabatic, \ie, $|B/\dot B| \gg \tgyr$). For second order Fermi
acceleration, both constraints, $\tacc<\Tinj$ and $\tacc<\tad$, are
equivalent, because the injection time is limited by the adiabatic drop of
the Alv\'en speed, which leads to $\Tinj \sim \tad$.

In conclusion, \eq{Enumax:Larmor} with $\cgyr{\,\lsim\,}\min(\frac13,
\thF^{-1})$ applies for our canonical assumption that the involved
hydrodynamic process are at least transrelativistic. Acceleration time
constraints then dominate for $\thF\gg 1$, which is particularly the case for
second order Fermi acceleration, or acceleration at nonrelativistic,
quasi-parallel shocks (\cf\ \RefApp{Fermi}), while for any faster
acceleration mechanism the geometrical extension of the emission region,
constrained by variability, sets the limit on $\Enuobs$.

\subsub{radcool}{Limits due to radiative cooling}

Synchrotron cooling of the protons during acceleration limits their maximum
energy through the condition $\tacc<\tsynp$. Writing the acceleration time as
$\tacc = 2\pi\thF\gamma_p/\omBp$ and using \Eq{phothad:tsyn}, we find
$\gamma_p < 3/\sqrt{8\pi \thF r_p \omBp}$, leading to
\eqn{Enumax:psyn}
\Enumaxobs \le \Enumaxobspsyn =  \frac3{8}\,\frac{m_\pi c^2 \cD}{r_p}
\sqrt{\frac{e}{2\pi\thF B}} \quad. 
\text
In the same way, proton acceleration must be faster than the cooling due to
photohadronic interactions, $\tacc < \tpg \approx \fpg^{-1}\tpi$. 
To express $\tpib$
by observable quantities, we use \eq{lum} to write
\eqn{tpib:lum}
\tpib = \frac{4\pi \cD^5 \cT^2 \rhoT^2}{\TauL} \with
\TauL = \frac{\Lbobs \Etanorm}{c^2\epsbobs\xL(a-1)^2}\quad,
\text
where $\Lbobs \equiv \cL(\epsbobs)$ is the isotropic luminosity of the source
at the observed spectral break, related to the break energy in the comoving
frame by $\epsbobs = \epsb\cD$, and $\Etanorm\approx 22\mubarn$ (\cf\ 
\RefApp{phothad}). Using \Eqs{phothad:tpi} we obtain from $\tacc<\tpg$
\eqn{Enumax:pg}
\Enumaxobs \le \Enumaxobspg = \frac{m_\pi c^2\gbt}4
	\left[\frac{\omBp \cT\rhoT}{\thF\UpsT\gbt}\right]^{\frac1a}
	\,\cD^{2+4/a} \quad
	\for \frac{2\omBp\cT^2\rhoT^2\cD^4}{\thF\fpg\TauL} \lsim \gbt\;,
\text
where we inserted the Doppler scaled characteristic Lorentz factor, $\gbt =
\eth/2\epsbobs$, which does not depend on $\cD$ and is related to the
comoving characteristic Lorentz factor by $\gb = \gbt\cD$, and
\eqn{UpsT}
\UpsT = \left\{\begin{array}{ll}
		\TauL\fpg/2\cT\rhoT 			&\for \gpmax\lsim\gb\\
		\TauL\fpg\Ngtot/2\Ngb\cT\rhoT      	&\for \gpmax \gg \gb
		\end{array}\right.\quad.
\text
The case of \eq{Enumax:pg} obviously corresponds to $\gpmax\lsim\gb$; the
case $\gpmax\gg\gb$ is described by setting $a=1$ and using the proper value
for $\UpsT$ [note: in \eq{tpib:lum}, the actual power law index $a$ has to be
used in any case]. It should be noted that \eq{Enumax:pg} only considers the
photon density in the burst connected to its intrinsic luminosity. If,
however, the relativistically moving ``blob'' is embedded in an ambient
photon field, which is isotropic with respect to the observer, the photon
density seen in the comoving frame of the blob can be considerably higher
than inferred from the observed luminosity by \eq{lum}. The reason is, that
this additional component would appear unboosted for the observer, and
therefore probably only as a small fraction of the apparent luminosity of the
emitting blob which is boosted by a factor $\cD^4$, while in the comoving
frame the photon number density of the {\em ambient} component is increased
by a factor $\cD$ due to Lorentz contraction and may thus dominate the photon
density. This scenario might be relevant in AGN, if the emission region in
the jet is close to the AGN core, and if the core radiation is isotropized by
a plasma halo (see, \eg, \cite{SBR94}), and photohadronic interactions may
limit the neutrino energy to values significantly below the upper limit
expressed by \eq{Enumax:pg}.

\sub{pimu}{Cooling of secondary particles in the hadronic cascade}

\subsub{cool}{Cooling processes and time scales}

Pions and muons are weakly decaying particles, with comparatively long
lifetimes, $\taupiRF = 2.6\mal 10^{-8}\scnd$ and $\taumuRF = 2.2\mal
10^{-6}\scnd$, respectively. For secondary particle Lorentz factors $\gsprod
\approx \gamma_p \gsim 10^{6{-}8}$, which are quite reasonable and readily
considered in most models, their lifetime in the comoving frame of the
emission region, $\tauscf = \tausRF\gs$, can be of the order the dynamical
time scale of the flare (\eg, in Gamma-Ray Bursts). Moreover, their
synchrotron losses are by a factor $(m_p/m_\decsym)^3 \sim 10^3$ stronger
than for protons. Adiabatic cooling of muons has been considered for neutrino
emission of Gamma-Ray Bursts \cite{WB97}, and synchrotron cooling of pions
and muons has been discussed for extremely magnetized environments, \eg,
neutron star magnetospheres \cite{RudM91b}. Most of the literature about
neutrino emission of AGN, however, neglects this effect. We will show here
that this is not justified, and derive the critical Lorentz factors, $\gsb$,
above which the energy loss of muons and pions plays a role. Obviously, we
have to distinguish between neutrinos from the decay of pions and muons,
because of their very different lifetimes. Energy losses of the muons are
generally more relevant, which affects in particular the electron neutrinos,
arising exclusively from muon decay. This is important for neutrino detection
in UHE air shower experiments, where electron neutrino showers are easier to
distinguish from the atmospheric background and are therefore proposed as
providing most of the expected signal \cite{PZ96,ZHV93}.

Secondary particles cool adiabatically prior to their decay if $\tauscf >
2|B/\dot B|$, which gives a critical Lorentz factor $\gsb = 2R/\alpha\bad
\tausRF$ for $B\propto R^{-\alpha}$, leading to
\eqn{Enumax:pimu:ad}
\Enumaxobssad = \Enudecs \;\frac{2\rhoT\cT\cD^2}{\alpha\bad\tausRF}\quad,
\text 
where $\Enudecs$ is the neutrino energy in the rest frame of the decaying
particle, which is roughly $\frac14 m_\pi c^2 \approx 35\MeV$ for pion decay,
but slightly smaller ($30\MeV$) for muon decay and may therefore be
distinguished (see \RefApp{phothad}). Analogously, pions and muons can
undergo efficient synchrotron cooling if $\tauscf \le \tsyns$. The critical
Lorentz factor is thus found from $\gb \tsynb (\mas/m_p)^3 = \gsb^2\tausRF$,
yielding
\eqn{Enumax:pimu:syn}
\Enumaxobsssyn = \frac{\omdecs}{\omBs}\;\Enudecs\cD \quad\with
\omdecs = \frac32\sqrt{\frac{c}{\tausRF \rs}}\quad, 
\text
where $\rs = e^2/\mas c^2$ is the classical radius of the particle, and $\omBs
= eB/\mas c$ its cyclotron frequency. The characteristic frequency, $\omdecs$,
for synchrotron losses of pions and muon is found as $\omdecpi =
5.0\mal10^{16}\scnd^{-1}$ and $\omdecmu = 4.7\mal 10^{15}\scnd^{-1}$,
respectively.  Secondary particles may also suffer inverse Compton (IC) losses
from interactions with background photons. The corresponding cooling time is
related to the synchrotron cooling time by the well known relation
\eqn{tics}
\tics = \tsyns \frac{\uph}{\uB} \approx \tsyns\;\frac{2 \Lb\bL}{B^2
\Rpar^2 c \xL}\quad,
\text
where $\bL$ is the bolometric correction factor relating the total photon
luminosity over the spectral range $\epsb$ to $\hat\eps$ to $\Lb$ given by
$\bL = |(\hat\eps/\epsb)^{2-a}-1|/|2-a|$ for $a\neq2$, and $\bL \approx
\ln(\hat\eps/\epsb)$ for $a=2$. For high pion and muon energies \eq{tics}
might be modified by Klein-Nishina corrections, leading to a suppression of
IC cooling.

Unlike the proton case, IC cooling is the most relevant process for
photo-interactions of pions and muons. Their IC cooling time is related to
the proton IC cooling time as $\tics = (\mas/m_p)^3 \ticp$, while the
Bethe-Heitler cooling time only scales with $\tBHs = (\mas/m_p) \tBHp$. For
$a=2$, one can show that $\tBHs \sim 5 \tics$, if Klein-Nishina corrections
are disregarded. For the pion, there are additional channels due to meson
resonance excitation; the lowest energy process is
$\pipm\gamma\to\rhopm\to\pipm\piO$, which has a theoretical peak cross
section of ${\sim}\,50\mybarn$ (determined from the $\rhopm\to\pipm\gamma$
decay branching ratio by use of the Breit-Wigner formula \cite{PDG96}), at a
photon energy of $\epspirho \sim 2\GeV$ in the pion rest frame. Compared to
the pion production off the proton via the $\mD$-resonance, where the
characteristic photon energy is $\epspD \sim 300\MeV$, this interaction is
suppressed by a factor $\sim \frac16(\epspD/\epspirho)^a$; for $a\approx 2$,
and neglecting the finite lifetime of the pion, secondary photon scattering
has a probability of ${<}\,0.3\%$, and can thus be neglected. However, the
process may be relevant in inverted photon spectra, \eg, for pion production
in the Rayleigh-Jeans part of the thermal background.

\subsub{specmod}{Spectral modification at the critical energies}

The ``maximum'' neutrino energies derived in \refsec{Epmax} arise from the
balance of energy gain and loss processes. The stochastic nature of these
processes allows the particles to exceed such ``limits'' with some, usually
exponentially decreasing probability. This has been shown for Fermi
accelerated particles subject to synchrotron losses \cite{WDB84/KA94}, which
show a largely unmodified extension of the power law spectrum up to the
cutoff energy (defined by balance of gains and losses), followed by an
exponential-like cutoff; depending on the detailed parameters, a pile-up may
occur at the cutoff energy. Although the stochastic behavior of photohadronic
losses is quite different from synchrotron losses, we may expect a similar
result in this case, and also for adiabatic losses. In general, we can assume
that if the neutrino energy is limited by the maximum proton energy, the
spectrum will continue as an approximate power law up to $\Enumaxobs$, and
drop off rapidly for $\Enuobs>\Enumaxobs$.

The situation is different for the decay of unstable particles, where the
particle can decay within a time $\Delta t$ after its production with a
probability $\Pdec = 1 - \exp(- \Delta t/\tauscf)$.  Far above the critical
Lorentz factor, $\gsdec \gg \gsb$, the decay probability within a cooling
time scale in the fluid frame, $\Delta t = \tcool(\gs)\ll \tauscf$, can be
approximated as $\Pdec(\tcool) \approx \tcool(\gs)/\tauscf$. The critical
Lorentz factor is defined by the condition that $\tcool(\gsb) = \gsb\tausRF$,
which allows us to write $\tcool(\gs) = \tauscf (\gs/\gsb)^{-w}$. For
adiabatic cooling, $\tcool = \const(\gs)$ then means $w=1$, while for
synchrotron cooling we have $w=2$; cooling by secondary photon scattering ---
if relevant --- would correspond to $w=a$.  For pions or muons produced with
$\gsprod \lsim \gsb$, neutrinos are produced with an energy $\Enuobs \approx
\Enudecs \gamma_p \cD$ in the observers frame, and their power spectrum is
$\LTnuobs\propto \Enuobs^q$, as discussed in \RefSec{phothad:specshape}. For
$\gsprod \gg \gsb$, the power spectrum is modified by the probability to
decay within a cooling time $\tcool(\gsprod)$, \viz,
\eqn{pimucool:mod}
\LTnuobs \propto \Enuobs^q\;\Pdec(\tcool)\Big|_{\gs =\Enuobs/\cD\Enudecs}
 \propto \Enuobs^{q-w} \for \Enuobs \gg \Enudecs \gsb \cD\;. 
\text
Therefore, the critical Lorentz factor marks a spectral break of magnitude
$\Delta q = -w$, rather than an exponential cutoff.  Clearly, this simplified
analytical estimate does not treat exactly the energy evolution of the pions
and muons, thus neglects particle number conservation. Considering this would
lead to a pile-up of decaying pions and muons around their respective
critical Lorentz factor, before the spectrum turns over into a
$\Enuobs^{q-w}$ behavior. The strength of the pile-up is correlated to the
magnitude of the spectral break, and is therefore expected to be stronger for
synchrotron cooling breaks than for adiabatic cooling breaks. The break due
to adiabatic cooling of secondary particles is comparable to the spectral
breaks occurring at the transition between different dominant proton cooling
processes, \Eq{phothad:specnu:q}. The $\Delta q=-2$ break caused by pion or
muon synchrotron cooling, however, is stronger and can be easily confused
with an exponential steepening: it causes a drop of events of one order of
magnitude over half a decade in energy, similar to the exponential function
around its critical energy. In practice, we may therefore consider
$\Enumaxobsssyn$ as a cutoff energy of the neutrino spectrum and compare it
with with the cutoff energies due to proton cooling as derived in
\refsec{Epmax}.

\sub{modind}{Model-independent discussion of spectral shapes, maximum
energies and fluxes} 

\subsub{parspace}{The parameter space}

The free parameters describing a transient fall into two classes: (i)
observable parameters, \ie, the characteristic time scale, $\cT$, and the
isotropized luminosity, $\cL$, and (ii) theoretical parameters, like $B$,
$\cD$, $\thF$, etc. We consider the former as given for any specific
transient (disregarding possible problems in their determination, \cf\
\refsec{times:intshock}), while the latter can only be constrained through
general physical considerations or additional observations within a certain
range. \eqs{Enumax:Larmor}, \refeq{Enumax:psyn}, and \refeq{Enumax:pg} show
that the maximum energies depend strongly on some of these parameters, in
particular on the magnetic field $B$ and the Doppler factor $\cD$. In
contrast, the hydrodynamic parameters of the flow, \ie, $\alpha$, $\bex$ and
$\bsh$ are of order unity, and can generally be recast into some reasonable
assumption for $\cgyr$. A special role is played by $\thF$, which describes
the speed of the acceleration process: as shown in \RefApp{Fermi}, first and
second order Fermi acceleration is limited to $\thF>1$, but we could easily
consider any faster acceleration process in our analysis by inserting the
appropriate $\thF$. We therefore shall use $\thF$ as a fixed parameter, with
the canonical assumption that $\thF\approx 1$, while $B$ and $\cD$ span the
two-dimensional parameter space describing a transient. From
\eqs{Enumax:Larmor}, \refeq{Enumax:psyn}, and \refeq{Enumax:pg} we can
immediately derive a qualitative division of the parameter space:

(1) For any given $\cD$, synchrotron losses of protons dominate over both
photohadronic or adiabatic (Larmor) limits for magnetic fields larger than
some value $\Bpsyn(\cD)$. Analogously, there is a magnetic field
$\Bssyn(\cD)$ above which muon or pion synchrotron losses dominate over all
other proton loss processes; the relation of $\Bpsyn(\cD)$ and $\Bssyn(\cD)$
depends on the other parameters.

(2) For any given $B$, photohadronic interactions of protons dominate over
both synchrotron and adiabatic (Larmor) limits for Doppler factors smaller
than some value $\Dpg(B)$. Similarly, there is a limit $\Dpgs$ below which
photohadronic interactions also dominate over pion or muon synchrotron
cooling. 

(3) For any given $\cD$, there is a limiting magnetic field $\Bpad(\cD)$
below which adiabatic or Larmor limits dominate over synchrotron or
photohadronic cooling of protons, and another value $\Bpads$ below which it
dominates over either muon or pion synchrotron cooling.

It is much more difficult to determine under which conditions adiabatic
cooling of secondary particles dominates; we will see in \RefSec{astro:GRB}
that, if at all, this can happen only in a very limited region of the
parameter space. For a more illustrative discussion of the parameter space
constraints specific to AGN jets and Gamma-Ray Bursts see
\RefSec{astro:AGN:spec} and \RefSec{astro:GRB:spec}, and the figures shown
there.

To get a quantitative idea about the above parameter space division, we
derive the condition under which the delimiting energies $\Enumaxobspsyn$,
$\Enumaxobspg$, and $\Enumaxobsgyr$ are all equal. This is equivalent to the
condition $\tacc = \tsynp = \tpg = 2\pi\rhoL\thF R/c$, representing three
equations which can be solved for the three variables $B$, $\cD$ and
$\gamma_p$; note that the latter term reduces to $2 R/c \alpha \bad = \tad$
in the case $\rhoL = (\pi\alpha\bad\thF)^{-1}$, \ie, that adiabatic cooling
rather than space limitations determines the Larmor limit. The corresponding
solutions for the magnetic field, Doppler factor, and maximum proton Lorentz
factor are called $\Bstar$, $\Dstar$, and $\gpstar$, respectively, and are for
the case $\gpmax\le\gb$ given as
\eqns{starpoint}
\subeq{D}  \Dstar &=& \Big[\UpsT^3\XiT^{a-1}\rhoL^{a+2}\thF^{4-a}\Big]^y\\
\subeq{B}  \Bstar &=& \Bb \Big[\UpsT\XiT^{a+3}\rhoL^{a+4}\thF^3\Big]^{-2y}\\
\subeq{gp} \gpstar &=& \gbt\Big[\UpsT\XiT^{a+3}\rhoL^{a+4}\thF^{-(a+2)}\Big]^y
\quad,
\text
with $y=(2a+10)^{-1}$, where we introduced the magnetic field scale $\Bb = 9
m_p^2 c^4/8 \pi e^3 \gbt^2 = [7.3\mal10^{21}\G]\gbt^{-2}$, and the
dimensionless quantity
\eqn{XiT}
\XiT\equiv\frac9{8\pi}\frac{c\cT\rhoT}{r_p\gbt^3}\quad.
\text
The relations for the case $\gpmax\gg\gb$ are obtained from
\eqs{starpoint} by setting $a=1$ and use the proper value of $\UpsT$ from
\eq{UpsT}. The corresponding maximum neutrino energy is obviously given as
$\Enuobsstar = \frac14 m_\pi c^2\Dstar\gpstar$. \eqs{starpoint} define a
unique reference point in the parameter space of magnetic field and Doppler
factor, which allows the discussion of the relation of cooling processes of
the proton independent of any other physical properties of the transient; for
notational simplicity, we refer to it as the {\em star-point} hereafter.

For some applications it is useful to normalize the magnetic field energy
density by the comoving photon energy density. We introduce the equipartition
parameter
\eqn{eqBg}
\eqBg \equiv \frac{\uB}{\uph} = \frac{c^3\rhoT^2\cT^2\xL B^2\cD^6}{2 \bL\Lbobs}
\quad,
\text
where $\bL$ is the bolometric correction factor of the photon spectrum as
defined in \eq{tics}. Without protons, $\eqBg\sim 1$ would correspond to
approximate energy equipartition of magnetic field and relativistic
particles, $\uB\sim \ue\sim\uph$, because of the high radiative efficiency of
electrons. In baryon loaded flows this is not generally the case because of
the contribution of relativistic protons with low radiative efficiency, and
$\up+\ue\sim\uB$ may imply $\eqBg\gg 1$ if the acceleration process works
more efficiently for protons. The corresponding star-point value is
\eqn{eqBgstar}
\eqBgstar = \frac{m_p c^3 \gbt^2\xL}{2\bL\Lbobs r_p}
	\Big[\UpsT^7\XiT^{3a+1}\rhoL^{a-2}\thF^{6-3a}\Big]^{2y}\quad.
\text

\subsub{uplim}{General upper limits on the neutrino energy}

For any given Doppler factor, the highest neutrino energy can be achieved for
$B = \Bpsyn(\cD)$, because with increasing $B$, $\Enumaxobsgyr$ increases and
$\Enumaxobspg$ remains constant, while $\Enumaxobspsyn$ decreases. The
equations determining $\Bpsyn$ are $\Enumaxobsgyr=\Enumaxobspsyn$ for
$\cD<\Dstar$ and $\Enumaxobspg=\Enumaxobspsyn$ for $\cD\ge\Dstar$, leading to
\eqns{Bpsyn}
\subeq{pg} \Bpsyn =& \mlbox{6}{\Bstar (\cD/\Dstar)^{-k}}  &\for \cD<\Dstar\\
\subeq{ad} \Bpsyn =& \mlbox{6}{\Bstar (\cD/\Dstar)^{-2/3}}&\for \cD\ge\Dstar
\quad,
\text
with $k=2+4/(a+2)$ in the case $\gpmax \lsim \gb$, and $k = \frac{10}3$ for
$\gpmax\gg \gb$. If we use $\eqBg$ as a coordinate of the parameter space
instead of $B$, and the maximum neutrino energy is reached for $\eqBgpsyn =
\eqBgstar (\cD/\Dstar)^{14/3}$ for $\cD>\Dstar$, and $\eqBgpsyn = \eqBgstar
(\cD/\Dstar)^{6-2k}$ otherwise. This provides an upper limit for the maximum
neutrino energy as a function of $\cD$
\eqns{Enumax:D}
\subeq{pg} \Enumaxobs \le \Enuobsplim(\cD) =& 
	\mlbox{6}{\Enuobsstar (\cD/\Dstar)^{1+k/2}}  &\for\cD < \Dstar\\
\subeq{ad} \Enumaxobs \le \Enuobsplim(\cD) =&
	 \mlbox{6}{\Enuobsstar (\cD/\Dstar)^{4/3}}   &\for\cD\ge\Dstar\quad.
\text
Stronger constraints on $\Enumaxobs$ may exist from secondary particle cooling. 
We will only consider synchrotron cooling, because the impact of adiabatic
cooling of secondary particles on the neutrino spectral shape is relatively
weak, as argued in \refsec{pimu:specmod}; moreover, it is often dominated by
the other cooling processes, as we will see in \RefSec{astro}.  From the
conditions $\Enumaxobsssyn=\Enumaxobspg$ for $\cD\lsim\Dstar$, and
$\Enumaxobsssyn=\Enumaxobsgyr$ otherwise, we find the solutions for the
magnetic field
\eqns{Bssyn:D}
\subeq{pg} \Bssyn(\cD) \approx& \mlbox{8}{\Bssyn^* (\cD/\Dstar)^{-\ks}} 
			&\for \cD\lsim\Dstar\\
\subeq{ad} \Bssyn(\cD) =& \mlbox{8}{\Bssyn^* (\cD/\Dstar)^{-1/2}} 
			&\for \cD\gsim\Dstar\;,
\text
where $\ks = 1+3/(a+1)$, and $\Bssyn^* = \sqrt{\BT\Bs/\rhoL\Dstar}$
with $\BT\equiv m_p c/e\rhoT\cT$ and $\Bs\equiv \mas\omdecs c/e$. Using
$\eqBg$ as a variable, we obtain $\eqBg \approx \eqBgssynstar
(\cD/\Dstar)^{6-2\ks}$ and $\eqBg = \eqBgssynstar (\cD/\Dstar)^5$,
respectively, with
\eqn{eqBgssynstar}
\eqBgssynstar = 
	\frac{c^3\rhoT^2\cT^2\xL\BT\Bs{\Dstar}^5}{2\rhoL\bL\Lbobs} \quad.
\text  
The transition from photohadronic to adiabatic dominance at $B=\Bssyn(\cD)$
is exactly at $\cD=\Dstar$ in the case $\gpmax\gg\gb$. For $\gpmax\le\gb$ the
exact transition value could be easily found by setting \eq{Bssyn:D:pg} and
\eq{Bssyn:D:ad} equal and solving for $\cD$, but the approximate division at
$\Dstar$ will usually be adequate in practice.  Secondary particle cooling
determines the maximum energy in the case $\Bssyn(\cD)<\Bpsyn(\cD)$, and the
corresponding upper limits to the neutrino energy are
\eqns{Enumax:D:pimu}
\subeq{pg} \Enumaxobs \le \Enuobsslim(\cD) = & \mlbox{7}{
		\Enuobsslim(\Dstar) (\cD/\Dstar)^{\ks+1}} &\for\cD\lsim\Dstar\\
\subeq{ad} \Enumaxobs \le \Enuobsslim(\cD) = & \mlbox{7}{
		\Enuobsslim(\Dstar) (\cD/\Dstar)^{3/2}}\quad&\for\cD\gsim\Dstar
\quad,
\text
with
\conteqno\eqns{Enumax:D:pimu}
\subeq{star} \Enuobsslim(\Dstar) 
		= \Enudecs\sqrt{\rhoL\rhoT\cT\mas\omdecs{\Dstar}^3/m_p}\quad.
\text
Obviously, $\Dstar$ can be replaced by any reference Doppler factor in
\eq{Enumax:D:pimu:ad}.  A more illustrative discussion of the relation
between $\Bssyn(\cD)$ and $\Bpsyn(\cD)$ in the different regions of the
parameter space is given in \RefSec{astro}.

\subsub{efficiencies}{Efficiency considerations and neutrino flux limits}

Relevant for the observability of ultra-high energy neutrinos is not only the
maximum neutrino energy a specific class of sources can provide, but also
which total neutrino luminosity is associated to this energy. This requires a
discussion of the neutrino efficiency as a function of the parameter space.

If we assume that $B<\Bssyn$, \ie, we are in the part of the parameter space
where secondary particle cooling plays no role, then the neutrino efficiency
can be written as $\Xnu = \frac12 \tpcool/\tpi$, which can be rewritten as
$\Xnu = \frac12\fmax^{-1}$ if one cooling process clearly dominates. In the
star-point, where all cooling times are equal at the maximum proton energy,
the neutrino efficiency at this energy is most easily derived as $\Xnustar =
(6\fpg^*)^{-1}$, with $\fpg^*\approx 3 + \exp(5a-10)$ since both terms for
neutron reconversion in \Eq{phothad:fpg} are small in the star-point. Using the
scaling properties of the cooling times, we can then derive the following
simple expressions:

If photohadronic cooling dominates,
\eqns{Xnu}
\subeq{pg} \Xnu \approx 
	\frac12\fpgmin^{-1} \with \fpgmin \approx 1 + \exp(5a-10)\;,
\text
which leads to $\Xnuad\approx \frac12$ if $a\le2$, \ie, if Bethe-Heitler
losses can be neglected. In general, the total hadronic radiative efficiency
in this region is ${\approx}\,1$, which approximately equal parts (for
$a\le2$) radiated in neutrino and electromagnetic channels.  Cosmic ray
ejection is generally suppressed by neutron reabsorption, and direct proton
ejection can usually assumed to be marginal, except for special geometries
and, maybe, at the highest energies.

If adiabatic cooling dominates,
\conteqno\eqns{Xnu} 
\subeq{ad} \Xnu \approx 
		3 \Xnustar \left(\frac{\cD}{\Dstar}\right)^{\!-(3+a)}
		\left(\frac{\gamma_p}{\gpstar}\right)^{\!a-1} \quad, 
\text
where $\Xnustar$ is the efficiency at the star-point. Here, most of the
energy is not radiated, but reconverted into kinetic energy of
expansion. Thus, the total hadronic radiative energy decreases in total,
while the distribution of the radiated energy between cosmic rays, neutrinos
and photons remains constant, and approximately $2:1:1$ for $a\le2$.

If synchrotron cooling dominates,
\conteqno\eqns{Xnu}
\subeq{syn} \Xnu \approx
		3 \Xnustar \left(\frac{\fpg^*}{\fpg}\right)
		\left(\frac{B}{\Bstar}\right)^{\!-2}
		\left(\frac{\cD}{\Dstar}\right)^{\!-(4+a)}  
		\left(\frac{\gamma_p}{\gpstar}\right)^{\!a-2}\quad.  
\text
Like in the case of photohadronic dominance, the total hadronic radiative
efficiency is close to 1, but the predominant part of the energy is emitted
in electromagnetic radiation. The correction factor $1\lsim\fpg^*/\fpg <
3$ considers that in the region of synchrotron dominance neutron reabsorption
may or may not play a role, depending on the ratio of the photohadronic
interaction time scale to the crossing time.

Note that, instead of the star-point, any point in or at the border of the
respective dominance region could be used as a reference point in \eqs{Xnu}.
If secondary particle cooling plays a role, the situation is similar: For
synchrotron cooling, the total hadronic radiative efficiency remains
constant, but more energy is channeled into energetic photons; for adiabatic
cooling, additional energy is reconverted in bulk kinetic energy. We do not
discuss the scaling properties for these cases; rather, we will use
$\Enumaxobsssyn$ as limiting energy and discuss the efficiencies only for
$\Enuobs<\Enumaxobsssyn$, and neglect the effect of adiabatic cooling of
secondary particles altogether.

The neutrino efficiency alone does not allow to derive flux rates; we also
have to make assumptions about the energy density of relativistic protons in
the source. Here we can use the standard equipartition argument discussed
above, and introduce a parameter $\eqpB\equiv \Upint/\uB \sim 1$, which
allows us to express the time integrated proton injection energy density,
$\Upint$ by the parameter space variable $\eqBg$ defined in \eq{eqBg}. The
neutrino luminosity at the maximum energy can then be expressed relative to
the bolometric photon luminosity as $\LRnu \equiv \LTnuobs/\bL\Lbobs\cT$, and
we define
\eqn{LRnumax}
\LRnulim \equiv \left.
	\frac{\LTnuobs}{\bL\Lbobs\cT}\right|_{\cE_\nu=\Enuobslim(\cD)} 
	= \left.\frac{\eqpB\eqBg\Xnu}{\bp}\right|_{\gamma_p=\gplim(\cD)}\quad, 
\text
where $\gplim$ denotes the maximum proton Lorentz factor attainable for a
given $\cD$ for which secondary particle cooling can yet be neglected.  If
$\Bssyn(\cD)>\Bpsyn(\cD)$, $\eqBg$ and $\Xnu(\gpmax)$ have to be evaluated at
$B=\Bpsyn(\cD)$ and we obtain
\eqns{LRnumax:psyn}
\subeq{pg} \LRnulimp =& \mlbox{12}{\frac{\eqpB\eqBgstar}{4\bp\fpgmin} 
			\left(\frac{\cD}{\Dstar}\right)^{\TS\frac{2a-4}{a+2}}} 
			& \for \cD<\Dstar\\
\subeq{ad} \LRnulimp =&  \mlbox{12}{\frac{\eqpB\eqBgstar}{4\bp\fpg^*} 
			\left(\frac{\cD}{\Dstar}\right)^{\TS\frac{4-2a}{3}}} 
			& \for \cD>\Dstar\;,
\text
while for $\cD=\Dstar$ we obviously have $\LRnulimp = \LRnustar \equiv
\eqpB\eqBgstar/6\bp\fpg^*$. If $\Bssyn(\cD)<\Bpsyn(\cD)$,
$\eqBg=\eqBgssyn(\cD)$ has to be used to determine $\gplim(\cD)$, leading to
\eqns{LRnumax:ssyn} 
\subeq{pg} \LRnulims \approx& \mlbox{12}{\frac{\eqpB\eqBgssynstar}{4\bp\fpgmin} 
			\left(\frac{\cD}{\Dstar}\right)^{\TS\frac{4a-2}{a+1}}} 
			& \for \cD<\Dstar\\
\subeq{ad} \LRnulims \approx&  \mlbox{12}{\frac{\eqpB\eqBgssynstar}{4\bp\fpg^*} 
			\left(\frac{\cD}{\Dstar}\right)^{\TS\frac{3-a}{2}}} 
			& \for \cD\ge\Dstar\;.
\text
We note that, if secondary particle cooling plays no role, $\LRnulim$ is
independent of $\cD$ for $a=2$, and has a maximum for $\cD=\Dstar$ for
$a>2$. In case of the dominance of secondary particle cooling, a maximum is
only obtained for $a\ge3$, while otherwise $\LRnulim$ continues to rise with
$\cD$. Therefore, for given $\Lbobs$, both the maximum neutrino energy and
the power flux at this energy increases with $\cD$ in most cases, and a
general upper limit cannot be stated. However, we have to consider that
increasing $\cD$, while maintaining $\gpmax=\gplim$, implies that also $\eqBg$
and therefore the total energy dissipated into relativistic particles and
magnetic fields by the transient, $\Etottrans$, increases. Since $\Etottrans$
is usually given, or at least limited by fundamental principles for any
specific kind of source model, we can impose an upper limit
\eqn{eqBgmax}
\eqBg < \eqBgmax \equiv \frac1{(1+\eqpB)}\left(\frac{4\pi}{\Omega}\,
		\frac{\Etottrans}{\bL\Lbobs\cT} - 1\right)\quad,
\text
where $\Omega$ is the total angle over which the energy is emitted; the case
of emission in a thin, freely expanding jet corresponds to $\Omega \approx
\cD^{-2}$.  This can be transformed into an upper limit on the Doppler factor
for which both $\Enumaxobs = \Enuobslim$ and $\LTnuobs(\Enumaxobs) = \LRnulim
\bL\Lbobs\cT$ can be attained, which we call $\Dlim$; for simplicity, we
confine the discussion to the case $\eqBgmax>\eqBgssynstar$, which means
$\Dlim >\Dstar$, and assume $\eqBgmax\gg 1$. From \eq{Enumax:D:pimu:ad} we
then obtain a general, model-independent upper limit on the neutrino energy
as
\eqns{Enuobsuplim}
\subeq{p}  \Enuobslim \le \Enuobsplim(\Dlim) =& 
\mlbox{18}{\Enuobsstar \left[
	\frac{4\pi\Etottrans}{\Omega^*\bL\Lbobs\cT(1+\eqpB)\eqBgstar}
	\right]^{\kappa_p}} &\for \Bssyn(\Dlim) \ge \Bpsyn(\Dlim) \\
\subeq{pimu} \Enuobslim \le \Enuobsslim(\Dlim) =& 
\mlbox{18}{\Enuobsslim(\Dstar)\left[
	\frac{4\pi\Etottrans}{\Omega^*\bL\Lbobs\cT(1+\eqpB)\eqBgssynstar}
	\right]^{\kappas}} &\for \Bssyn(\Dlim) < \Bpsyn(\Dlim)\;,
\text
where $\Enuobsslim(\Dlim)$ is given by \eq{Enumax:D:pimu:star}. The power
law indices in \eqs{Enuobsuplim} are: $\kappa_p=\frac27$ and $\kappas =
\frac3{10}$ for $\Omega^*=\Omega=\const > \Dlim^{-2}$, while $\kappa_p =
\kappas = \frac12$ with $\Omega^*={\Dstar}^{-2}$ in the case of a free jet,
$\Omega\sim \cD^{-2}$. \eq{Enuobsuplim} is indeed a true upper limit for
$\Enumaxobs$: increasing $\cD$ beyond $\Dlim$ while keeping $\eqBg=\eqBgmax$,
implies $B\propto \cD^{-3}$ for constant $\Omega$, and $B\propto \cD^{-2}$
for a free jet; since the maximum energy is determined by the adiabatic
limit, \eq{Enumax:Larmor}, this leads to $\Enumaxobs \propto \cD^{-1}$ and
$\Enumaxobs\propto\const$, respectively. It is obvious that any other choice
of parameters, \eg, $\cD<\Dlim$ or $\eqBg<\eqBgmax$, leads to lower limiting
energies. From \eq{LRnumax:ssyn:ad} we obtain
\eqns{LRnulim:Dlim}
\LRnulim \le \LRnulimp(\Dlim) =& 
	\mlbox{18}{\frac{\eqpB\eqBgstar}{4\bp\fpg^*} 
	 \left[	\frac{4\pi\Etottrans}{\Omega^*\bL\Lbobs\cT(1+\eqpB)\eqBgstar}
		\right]^{\lambda_p}} &\for \Bssyn(\Dlim) \ge \Bpsyn(\Dlim) \\
\LRnulim \le \LRnulims(\Dlim) =&
	\mlbox{18}{\frac{\eqpB\eqBgssynstar}{4\bp\fpg^*}  
	 \left[	\frac{4\pi\Etottrans}{\Omega^*\bL\Lbobs\cT(1+\eqpB)\eqBgstar}
		\right]^{\lambdas}} &\for \Bssyn(\Dlim) < \Bpsyn(\Dlim)\;, 
\text
with $\lambda_p = (2-a)/7$ and $\lambdas=(3-a)/10$ for
$\Omega=\Omega^*=\const$, and $\lambda_p = (2-a)/4$, $\lambdas=(3-a)/6$, and
$\Omega^*={\Dstar}^{-2}$ in the case of a free jet. \eq{LRnumax} is not an
upper limit, because higher values are generally allowed for lower Doppler
factors; thus, one can increase the emitted neutrino power compared to the
value in \eq{LRnulim:Dlim} on the expense of the maximum neutrino energy. We
also emphasize the role of the baryonic energy content: increasing $\eqpB$
increases $\LRnulim(\Dlim)$, but decreases $\Dlim$ and therefore the upper
limit on $\Enuobslim$. We note that the dependence on the value of $\Dlim$ is
low, so that in many cases $\Enuobsstar$ and $\LRnustar$ will give good
order-of-magnitude estimates on the possible ultra-high energy neutrino of a
transient. We illustrate this discussion in some more specific applications
in the next section.

\sect{astro}{Neutrino emission from unsteady astrophysical sources} 

In this section, we apply the general theory developed in \RefSec{theory} to
specific models of astrophysical transients, which are commonly discussed in
the literature. Our aim is to explore the parameter space of these models
more extensively than usually done in the literature, and to check the
results for consistency with the limits set by secondary particle cooling,
which is disregarded in most papers. To simplify the discussion, we ignore
in the following the geometric parameters $\xL$ and $\rhoT$ introduced in the
last section, and assume $\xL=\rhoT=1$. However, since all times are
normalized by the size of the emission region, and all luminosities by the
photon energy density in the comoving frame, they can easily be re-introduced
by replacing $\cT\to \cT \rhoT$ and $\cL\to \cL/\xL$ in all equations of this
section.

\sub{AGN}{Neutrinos from AGN jets}

There are two classes of AGN models which predict neutrino emission, both
involving normal $e$-$p$ flows. One class assumes particle acceleration at
shocks in the accretion flow very close to the black hole, and produce
neutrinos by both $p\gamma$ and $pp$ interactions
\cite{SDSS91,SP94,SS96,PS92}. The other applies to radio loud AGN, which show
extended radio jets, and locates the emission region at internal shocks in
the relativistic jets at larger distances from from the black hole
\cite{Man95app,SS96}. The highest energy neutrinos would then be expected
from {\em blazars}, which are AGN jets pointing in the direction of the
observer, because the energy is boosted by the Doppler factor $\cD \sim \Gjet
\gg 1$ \cite{MSB92}. We discuss this class of models, \ie, the {\em AGN jet
models}, in the following; some interesting implications of our results on
the other class, the {\em AGN core models}, are described in
\RefApp{AGNcore}.

\subsub{blazar}{The ``proton blazar'' scenario}

Blazars are known to emit electromagnetic radiation from radio wavelengths up
to the TeV-gamma-ray regime. Their spectrum shows a typical ``two-hump''
structure, where characteristic photon energies depend on the source
luminosity: In high luminosity blazars, such as 3C 279, the lower hump cuts
off at optical wavelengths, while the high energy emission extends up to at
least $10\GeV$; in low luminosity sources, like in the nearby objects
Mkn\,421 and Mkn\,501, the lower hump extends in flares up to $10{-}100\keV$
\cite{CBB+97,PVT+98}, and the high energy emission is observed up to
${\gsim}\,10\TeV$ \cite{ICRC97-TeV,AAB+97}. While there is agreement that the
low energy hump is due to synchrotron emission of energetic electrons, the
origin of the high energy emission is unclear: It can be explained (a) by
inverse-Compton emission of the same electron population producing the low
energy synchrotron radiation (\eg, \cite{SMB97,MK97}, and references
therein), which could arise also if the jets are leptonic (consisting of
$e^\pm$ and magnetic fields, with few or no protons); or (b) by
electromagnetic cascades induced from the decay of photohadronic pions
\cite{MKB91}. The latter mechanism is referred to as the ``proton blazar'' or
simply ``hadronic'' scenario, and gives rise to considerable neutrino fluxes
\cite{Man95app,Man93/96}, while the leptonic models obviously do not.  On the
basis of gamma-ray observations alone, the issue of the dominant radiation
process in AGN jets is not settled yet \cite{Buc/Man98sci}; the observation
of correlated neutrinos could resolve this issue.

The typical bulk Lorentz factors $\Gjet$ of AGN jets can be estimated from
the apparent superluminal motion of blobs in the jet; a recent investigation
of 43 AGN indicates that $\Gjet\lsim 30$ \cite{GD97}, and the typical
inclination angle of the blazar jets to the line of sight is inferred to be
$\exval{\phiview}_{\rm bl} \sim 5\deg$, confirming the estimate $\cD \sim
\Gjet \sim 10$ obtained from AGN unification models \cite{UPad95}; we will
use $\cD\lsim 30$ as an upper limit estimate. A lower limit on the Doppler
factor of TeV blazars can be found from the observed emission of photons with
$\veps\sim 1\TeV$, where the emission region must be optically thin with
respect to $\gamma\gamma\to\ep\ee$ reactions of gamma-rays on intrinsic
soft photons, which leads to $\cD \gsim 2 \LXAGN^{1/5} \cT_4^{-1/5}$, or
roughly $\cD\gsim 3$. A similar limit is obtained for the high luminosity
blazar 3C\,279, using EGRET observations and assuming the emission to be
optically thin at $1\GeV$ \cite{SBR94}.

In contrast to the bulk Lorentz factor, the magnetic field strength in blazars 
is more difficult to estimate, although hadronic models typically invoke $B\gsim
10\G$ based on equipartition arguments \cite{Man96heid}. A test of this claim
is possible by observing the synchrotron-self absorption frequency of the
{\em variable} emission of blazars; an analysis of the spectral shape and the
multifrequency variability of blazars suggests that this turnover is at an
observed frequency ${\sim}300\GHz$ \cite{BRG+89III}. Boosting into the
comoving frame of the blob with $\cD\lsim 10$, this would be consistent with
synchrotron-self absorption for $B\sim 10\G$ and a relativistic electron
energy density $\ue \ll B^2/8\pi$, which is expected in hadronic models
\cite{Man93/96}. 

The emission from blazars is strongly variable, with activity periods taking
turns with quiescent periods on a typical time scales of months
\cite{GMU+95}; this kind of long-term variability is observed at all
frequencies, and appears to be largely correlated. Also within an activity
period, the TeV emission from Mkn\,421 and Mkn\,501 shows clearly separated
flares with doubling time scales from days down to less than one hour
\cite{AAB+97,GAB+96}, \viz, $\cT\sim 10^3{-}10^6\scnd$, with correlated
variability of the synchrotron emission at X-ray energies \cite{BAB+96}. This
suggests an identification of these short-term flares with transient,
causally disconnected acceleration regions of energetic particles, \eg, as
expected in the scenario of internal shocks in the jets. 

\subsub{spec}{The parameter space for time-integrated neutrino spectra from
blazar flares}

We assume that the relevant target photons for photohadronic pion production
are the synchrotron photons in the low energy spectral hump produced by
accelerated electrons, since the number density of photons in the high energy
hump is too low and can be neglected. We confine the discussion to low
luminosity TeV blazars, where the low energy photon spectrum extends to
$\eps\gsim 1\keV$; for high luminosity blazars, like 3C\,279, in which the
synchrotron component cuts off at optical frequencies, our power law
approximation is not applicable and a more detailed calculation would be
required. The target photon spectrum in the comoving frame of the
relativistic flow can then be approximated by a power law with a typical
index $\exval{a} \approx 1.7$, if we use a power law interpolation between
the sub-mm and X-ray wave bands \cite{BRG+89III}, ignoring the observed break
at optical frequencies; this is also justified by the observation that in
flares the important optical-to-X-ray spectrum of low luminosity blazars
seems to be flatter than the typical $\exval{\aoX} \approx 2$ seen in the
quiescent state \cite{PVT+98,BAB+96,LW98}. In the following, we adopt
$a=\frac53$ and introduce the Doppler factor and magnetic field in canonical
units, $\cD_1\equiv \cD/10$, and $B_1 = B/10\G$. For the break energy we use
$\epsbobs \sim 10^{-4}\eV$ \cite{BRG+89III}, and $\hat\veps \sim 1{-}10\keV$
from X-ray observations of flares in Mkn\,421 and Mkn\,501
\cite{PVT+98,TTM+96}, leading to $\bL\sim 10^3$. The luminosity at the
spectral break is not taken from observed fluxes at $\epsbobs$, because the
emission at low energy is likely to be superposed by the emission from other
jet regions not associated with the flare; rather, we use the observed,
isotropized X-ray luminosity at $\epsXobs\approx 1\keV$ of the flare, $\LXobs
= [10^{45}\erg/\!\scnd] \LXAGN$, and determine $\Lbobs$ from scaling with the
assumed power law photon spectrum, $\Lbobs = \LXAGN
(\epsbobs/\epsXobs)^{1/3}$; note that $\bL\Lbobs \sim 3\LXobs$. Opacity
requirements suggest that for low luminosity TeV blazars the external
radiation does not dominate over the synchrotron component in the comoving
frame of the flow \cite{CFR98}; if this would be the case, the thermal-like
properties of the disk radiation would yield a neutrino spectra substantially
different from our results.

For the characteristic Lorentz factor in the comoving frame we find $\gb \sim
[3\mal10^{11}]\cD$, which is of the order the maximum proton Lorentz factors
observed in cosmic rays. We will show below that acceleration of protons in
blazars cannot reach higher Lorentz factors, and confine ourselves to the
case $\gamma_p < \gb$.  In order to ignore the upper limit in the photon
spectrum, we have to require $\gamma_p\gsim 10^{6{-}7}\cD_1$.

To estimate the relevant time scale for the transient emission, $\cT$, we use
the X-ray variability, where we have to consider {\em electron} cooling times
of order $[30\scnd]\,B_1^{-3/2} \cD_1^{-1}$ in the observer frame, which for 
standard parameters are much shorter than the observed rise or decay times of
the flare. This suggests that one can use \Eq{theory:Tcr}, which
explains the generally longer decay times by Lorentz boosting effects in
transrelativistic internal shocks, rather than by slow cooling. We note that
this is in contrast to the usual interpretation of $\cT$ as the doubling time
of the flare, which assumes much longer cooling times expected in the weak
magnetic fields required by purely leptonic emission models to explain
$\Tdecay>\Trise$. The latter explanation may also apply to hadronic
models, if second order Fermi acceleration plays a role, so that in principle
$\cT=\Tdecay-\Trise$ can only be considered as an upper limit. The typical
time scales of blazar flares are then $\cT \equiv [10^4\scnd]\cT_4$ with
$0.1\lsim\cT_4\lsim 10$, corresponding to a comoving linear size of the
emission region $\Rpar\sim [3\mal10^{15}\cm]\,\cT_4\cD_1$. The fact that we
probably have to deal with transrelativistic shocks also suggests 
$\thF\approx 1$, which we assume in numerical estimates; in general, however,
we keep $\thF$ as a free parameter.

In a free relativistic jet we have $\tad \approx \Tcr$, \ie, the emission
region expands with the velocity of light, but $\bad<1$ is also possible if
the jets are confined. Heuristically, we can express the value of $\bad$ by the
opening angle of the jet, $\bad \approx \min(1,\phiopen\Gjet)$, and assume
$\phiopen\lsim \exval{\phiview}_{\rm bl} \sim 0.1$, which means that $\Gjet
\sim 10$ corresponds to $\bad\lsim 1$.  Hence, existing observations cannot
decide whether magnetic confinement applies or not, but $\bex$ can be
expected to be close to $1$. On the other hand standard jet theory assumes
that $B\propto R^{-1}$ \cite{BRK-jet}, \ie, $\alpha=1$, but $\alpha=2$ may
apply if reconnection isotropizes the magnetic field as assumed in Gamma-Ray
Bursts \cite{CThom94}. Since $\bex$ and $\alpha$ always appear as a product
in the equation, $\alpha\bex\approx1$ is a reasonable assumption, and will be
used in the following. The energy limit set by adiabatic cooling then
corresponds to $\cgyr \approx \frac13\thF^{-1}$, so that our canonical
assumption $\thF\approx 1$ is equivalent to the assumption that particles are
accelerated up to their Larmor limit.
\ifsubmit\relax
\else
\setfig{tbp}{AGN:parspace}
\epsfxsize=0.95\textwidth\epsfbox[0 365 593 650]{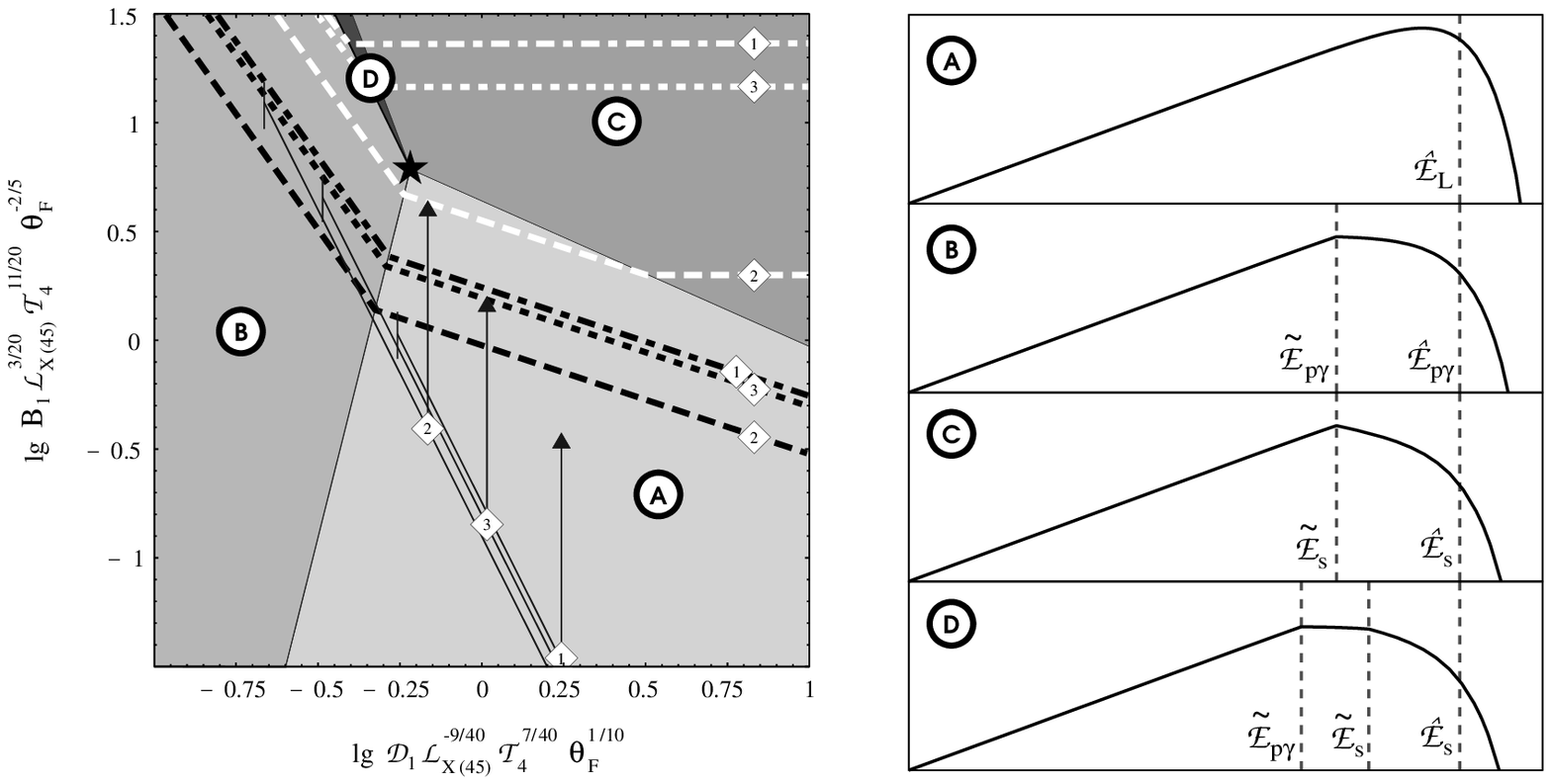} 
\figcap{\capsize Dominant cooling processes and neutrino spectral shapes for
AGN jets --- {\bf Left:} Parameter space, with the ``star-point'', denoting
equal cooling time scales at the maximum energy, indicated. The shaded
regions correspond to the dominant cooling process at the maximum proton
energy: (A) Larmor limit or adiabatic cooling; (B) photohadronic cooling; (C)
and (D) synchrotron cooling, where (D) marks the region where photohadronic
cooling dominates for a part of the energy spectrum. Also shown are the
positions of three observed AGN flares: $\exval{1}$ Mkn 421, April 26, 1995 
($\cT_4=10$, $\LXAGN=0.5$) \cite{BAB+96}; $\exval{2}$ Mkn 421, May 7, 1996
($\cT_4=0.1$, $\LXAGN=0.9$) \cite{Schub97}; $\exval{3}$ Mkn 501, April 16,
1997 ($\cT_4=3$, $\LXAGN=2.0$) \cite{PVT+98}. Central positions assume $\uB =
\uph$, black triangles correspond to $\uB=100 \uph$, diagonal errors indicate
the range of possible Doppler factors (see text). Numbers in diamonds
assiciate data points to the corresponding delimiting lines of muon cooling
(black) and pion cooling (white); secondary particle cooling is relevant in
the parameter space region above these lines. {\bf Right:} Schematic
representation of the shapes of neutrino spectra (time integrated power per
logarithmic interval of energy), $\ln \LTnuobs(\Enuobs)$ vs. $\ln \Enuobs$,
corresponding to regions (A)--(D). Break energies due to changes of the
dominant proton cooling process are indicated (\cf\ \RefSec{phothad:pcool}),
possible additional breaks due to secondary particle cooling are omitted for
simplicity (\cf\ \RefSec{theory:pimu:specmod}).}
\text\fi
Using these standard parameters, we find for the star-point of the parameter
space 
\eqns{AGN:starpoint}
\subeq{D}   \Dstar_1    \approx 	&\mrbox{2}{0.6} 		
			&\times\; \LXAGN^{9/40} \cT_4^{-7/40} \thF^{-1/10}\\
\subeq{B}   \Bstar_1    \approx 	&\mrbox{2}{6}
			&\times\; \LXAGN^{-3/20}\cT_4^{-11/20}\thF^{2/5}\\ 
\subeq{eqBg}\eqBgstar  \sim	&\mrbox{2}{50}
			&\times\; \LXAGN^{1/20}\cT_4^{-3/20}\thF^{1/5}\quad,
\text
corresponding to a maximum neutrino energy
\conteqno\eqns{AGN:starpoint}
\subeq{Enu} \Enuobsstar &\approx& 2\mal10^{18}\eV
		    	\;\times\; \LXAGN^{3/10}\cT_4^{1/10}\thF^{-4/5}\quad.
\text
\fig{AGN:parspace} shows the different regions of dominant cooling at the
maximum energies, and their associated spectral shapes, where we have scaled
$\cD_1$ and $B_1$ relative to the star-point values. Also shown are the
positions of three observed AGN flares in the parameter space, for which we
assumed the magnetic field to be in equipartition (a) with the radiation
density, or (b) with the energy density of protons, $\up=100 \ue$,
corresponding to standard assumptions in hadronic blazar models, and a range
of possible Doppler factors $0.3 \lsim\cD_1\lsim 3$. Comparison with the
regions of dominance of muon and pion cooling corresponding to these flares
shows that muon particle cooling plays a role for short flares in scenario
(b), if $\cD_1\lsim 1$; pion cooling is mostly unimportant for usual hadronic
AGN models. In most cases, the neutrino energy is limited by Larmor radius
constraints of the accelerated protons, consistent with earlier assumptions
\cite{HZ97,Man93/96}.

\subsub{Emax}{Blazar neutrino maximum energies and fluxes}

Keeping the magnetic field, the Doppler factor and the proton-to-electron
energy ratio in blazars as free parameters, rather than adopting common
assumptions, we can apply our discussion in \RefSec{theory:modind} to obtain
general upper limits on neutrino energies and fluxes from this class of
objects. We see from \fig{AGN:parspace} that, for parameters typically
observed in blazar flares, that $\Bmusyn(\cD)<\Bpsyn(\cD)$, but $\Bpisyn(\cD)
\gsim \Bpsyn(\cD)$ for all Doppler factors in the discussed range. Therefore,
$\Enuobs<\Enuobsplim(\cD)$ applies to muon neutrinos from pion decay, and
$\Enuobs<\Enuobsmulim(\cD)$ to muon and electron neutrinos from muon decay. 

To find out the relevant range for the Doppler factors, we start with
energetical considerations. The usual limit applied to the power of AGN jets
is the Eddington luminosity of the putative Black Hole in the AGN, 
$\Ljet\lsim [10^{47} \erg/\!\scnd] \MBH$, where $\MBH$ is the mass of the
Black Hole in units of $10^9\Msol$. Since we consider a beamed emitter, we
have to set $\Omega=\cD^{-2}$. Inserting in \Eq{theory:eqBgmax}, assuming an
energy dissipation efficiency of $\Xish\sim 0.2$ in the jet as expected for
transrelativistic internal shocks, and equipartition of proton and magnetic
field energy density, $\eqpB=1$, we obtain $\eqBgmax \sim [4\mal10^3]\,\MBH
\LXAGN^{-1}\cD_1^2$, thus for the limiting Doppler factor allowing
$\Enumaxobspi = \Enuobspilim$,
\eqn{Dlim:AGN:pi}
\Dlim_{1,\pi} \approx 2 \times \MBH^{3/8} \cT_4^{-1/4} \thF^{-1/4}\quad.
\text
For the canonical range of assumed AGN Black Hole masses, $0.1\lsim \MBH\lsim
10$, we can therefore assume that $\Dlim_\pi > \Dstar$, for which case we find
\eqns{AGN:limits:pi}
\subeq{E}  \Enuobspilim(\cD_1) \approx& \mrbox{12}{4\mal 10^{18}\eV
		\times D_1^{4/3} \cT_4^{1/3} \thF^{-2/3}} 
		& \lsim 1\mal 10^{19}\eV \times \MBH^{1/2} \thF^{-1} \\
\subeq{LR} \LRnulimpi(\cD_1) \approx& \mrbox{12}{
		0.3 \times D_1^{2/9} \cT_4^{-1/9} \thF^{2/9}}
		& \lsim 0.3 \times \MBH^{1/12} \cT_4^{-1/6} \thF^{1/6}\;,
\text
where we assumed $\bp=20$. Adopting these parameters, neutrinos from muon
decay are limited by muon synchrotron cooling to an energy $\Enumaxobsmu
\lsim [8\mal10^{17}\eV]\;\MBH^{5/8}\cT_4^{1/4}\thF^{-3/4}$.  On the other
hand, we can also find the conditions under which neutrinos from muon decay,
in particular electron neutrinos, can reach their highest energies and
fluxes. We determine $\eqBgmusynstar\approx
3\,\LXAGN^{1/8}\cT_4^{1/8}\thF^{1/2}$, leading to
\eqn{Dlim:AGN:mu}
\Dlim_{1,\mu} \approx 5 \times \MBH^{1/3} \cT_4^{-1/3} \thF^{-1/3} \quad.
\text
Again we can confine the discussion to the case $\Dlim_\mu>\Dstar$, which
gives
\eqns{AGN:limits:mu}
\subeq{E}  \Enuobsmulim(\cD_1) \approx& \mrbox{12}{1\mal 10^{18}\eV
		\times D_1^{3/2} \cT_4^{1/2} \thF^{-1/2}} 
		& \lsim 1\mal10^{19}\eV \times \MBH^{1/2} \thF^{-1} \\
\subeq{LR} \LRnulimmu(\cD_1) \approx& \mrbox{12}{2\mal10^{-2} 
		\times D_1^{2/3} \cT_4^{1/4} \thF^{3/5\phantom{-}}}
		& \lsim 5\mal10^{-2} \times \MBH^{2/9} \thF^{1/3}\;.
\text
The result that the upper energy limits for neutrinos from pion and muon
decay are equal in their respective optimization is a consequence of the
result that the maximum energy of pion neutrinos for Doppler factors
$\cD>\Dlim_\pi$ remains unchanged at the value $\Enuobspilim$, as discussed
in \RefSec{theory:modind:uplim}. For the assumed range of AGN Black Hole
masses and $\thF\gsim1$, we therefore obtain a {\em strict} upper limit of
about $3\mal10^{19}\eV$ for neutrinos from AGN flares, which is independent
of the flare time scale and any model assumptions. The inverse linear
dependence on $\thF$, however, shows that this energy limit would strongly
increase if we assume acceleration on time scales much shorter than the
particles Larmor motion.

If the TeV emission from blazars, which has generally a luminosity comparable
to the X-ray emission, ought to be explained by hadronic emission, the
corresponding neutrino luminosities would have to be of the same order, or
$\LRnulim\sim\frac13$. \eqs{AGN:limits:mu} show that this is incompatible
with the conditions neutrinos from muon decay (electron neutrinos) need to
reach their theoretical energy limit of ${\sim}10^{19}\eV$. For muon neutrinos
from pion decay, however, this seems to be possible for very large baryonic
and magnetic energy densities, $\eqBg\sim 10^3$; this scenario would expect a
difference between the electron and muon neutrino cutoff energies of more
than one order of magnitude.  It should be noted, however, that proton
blazars can produce $\LTnuobs(\Enumaxobs)\gsim\LXobs$ also for relatively
moderate values of $\eqBg$, if $\cD\approx \Dstar$; then, the hadronic
radiative efficiency (neutrinos + gamma-rays) increases to ${\gsim}50\%$, so
that a comparable emission in neutrinos, high energy and low energy photons
can be achieved for $\eqpB\sim\eqBg\sim 1$. This more realistic scenario
leads to maximum neutrino energies much below the upper limits stated in
\eqs{AGN:limits:pi:E} and \refeq{AGN:limits:mu:E}, so that flares from AGN
jets would not be expected to emit considerable neutrino fluxes above a few
times $10^{18}\eV$.

To get an estimate on event rates in current or planned VHE/UHE neutrino
observatories, we consider the example of the May 7, 1997 flare of Mkn 501,
which lasted $3\mal10^4\scnd$: The total isotropized energy emitted in
optical to X-ray photons of this flare is $\LTOX\sim 2\mal10^{50}\erg$;
taking the luminosity distance of ${\approx}160\Mpc$, and assuming
$\LTnuobs\sim \LTOX$ as suggested by TeV observations, this corresponds to a
total energy in neutrinos of $\Enumaxobs \sim 10^{18}\eV$ at earth of about
$6\mal 10^5\erg \km^{-2}$, which would produce ${\sim}3\mal10^{-7}$ neutrino
induced showers per $\km^3$ air volume. The biggest air fluorescence
detectors currently planned would cover about $3\mal 10^5\km^3$ air, and the
ground array of the Pierre Auger Observatory would correspond to about $5\mal
10^3\km^3$, which would clearly be not sufficient to detect the neutrino
emission of a single AGN flare. Following an $\Enuobs^{2/3}$ power spectrum,
the total energy in neutrinos at $\Enuobs\sim 10^{15}\eV$ would be about
$6\mal 10^3\erg\km^{-2}$, which would cause ${\sim}3\mal 10^{-3}$ events in a
$1\km^3$ underwater/ice Cherenkov detector. Also here, even the biggest
neutrino telescope currently considered would not be able to ``see'' single
AGN flares. The best we can expect is therefore to collect diffuse fluxes
corresponding to many AGN flares and determine the average properties of
their neutrino spectra. Details of the time integrated emission spectra of
AGN correlated transients, however, would still be important to determine
reliable estimates for such diffuse fluxes.

\sub{GRB}{Gamma-Ray Bursts}

Gamma-Ray Bursts (GRBs) are thought to be produced in highly relativistic
outflows originating from a compact, explosive event over time scales of less
than a second up to several minutes \cite{FM95ARAA,Mes98hunt}.  Recent
observations of GRB afterglows \cite{Kou97} indicate that they are
located at cosmological distances, which requires a characteristic luminosity
of about ${\gsim}10^{51}\erg\scnd^{-1}$ under the assumption of isotropic
emission. Possible scenarios which could release that much energy over the
time scales observed are, \eg, the coalescence of a neutron star binary, or
the collapse of a supermassive star. The radiation observed from GRBs is
expected to be due mainly to synchrotron or inverse Compton radiation from
relativistic electrons accelerated at shock fronts occurring near the
interface of the expanding relativistic shell ({\em external shocks}), or at
shocks forming within the unsteady outflow itself ({\em internal shocks})
\cite{Mes95texas}. The same mechanisms would also accelerate protons, which
could reach energies of the order of the highest energy cosmic rays, ${\sim}
10^{20}\eV$ \cite{Wax95,Vie95}. Due to the interaction with the dense photon
field in the burst, these protons can produce efficiently VHE, and maybe also
UHE neutrinos \cite{WB97}. Neutrinos of lower energy may also be produced by
$pp$ interactions between cold protons in the colliding ejecta \cite{PX94}.
Obviously, both scenarios fall into the class of transient emission, and we can
apply our results from \RefSec{theory} to examine in more detail the spectrum
and maximum energy of the neutrinos from $p\gamma$ interactions in Gamma-Ray
Bursts. We note, however, that our discussion assumes that the physical
parameters in the transient remain approximately constant over the emission
time scale; it does therefore not apply to GRB afterglows, in which the
parameters change drastically over very long time scales.

\subsub{fireball}{Cosmological fireball models and internal shocks}

Many GRBs show intrinsic variability on time scales $\cT \sim 1\ms{-}1\scnd$,
while the total burst durations are typically $\Tburst \sim 0.1{-}100\scnd$
\cite{FM95ARAA}. This implies that their energy is released in a volume of
the typical dimension of compact or stellar objects, $R_0 \sim
10^7{-}10^{10}\cm$. For a total energy of ${\gsim}10^{52}\erg$, this leads to
a local photon density of ${\gsim}10^{21}\erg\cm^{-3}$ with photon energies
${\gg}m_e c^2$. These ``fireballs'' (and even much weaker ones as well) would
be optically thick to $\gamma\gamma$ pair creation, and for small baryonic
loads, $\Mbar \gsim 10^{-6}\Msol$, the expansion leads to a conversion of
almost all the radiation energy into bulk kinetic energy of motion,
accelerating to a limiting Lorentz factor of $\Gamma \sim \eta =
\Eburst/\Mbar c^2 \gg 1$, before photons can escape \cite{MRees93apj}. Hence, a
dissipation mechanism reconverting the bulk kinetic motion into radiation is
required after the flow becomes optically thin: this can be achieved by
electron acceleration at shock waves occurring when the ejecta run into
external matter \cite{MRees93apj}. Moreover, internal shocks can form in the
ejected wind if the outflow is non-steady, \ie, $\eta$ varies significantly
over time scales ${\ll}\Tburst$, which can lead to faster shells catching up
with slower shells \cite{ReesM94}, similarly to what was discussed for AGN in
\RefSec{theory:times}. Such internal shocks can dissipate the kinetic energy
with an efficiency comparable to external shocks. While external shocks are
expected to produce a relatively featureless outburst over time scales
comparable to the total burst duration, $\Tburst$, internal shocks could be
associated with the rapid variability within the burst on time scales
$\cT\ll\Tburst$. In both external and internal shocks a substantial fraction
of the gamma radiation is produced by synchrotron cooling of the
shock-accelerated electrons. At the dissipation radius where internal shocks
occur \cite{ReesM94},
\eqn{GRB:rd}
\rd \sim c \cT \eta^2 \quad,
\text
\ie, where the radiated gamma-rays are produced, the comoving magnetic field
energy density can be parametrized through $\uB = \eqBg \uph$, so that
\eqn{GRB:Beq}
B \sim 10^{10}\G\times (\eqBg \bL \LGRBobs)^{1/2} \TGRBobs^{-1} \eta^{-3}\quad,
\text
where $\TGRBobs = \cT/1\scnd$ and $\LGRBobs = \cL/10^{51}\erg\scnd^{-1}$ are
the normalized GRB variability time scale and isotropic luminosity at the
break energy in the observer frame, respectively, and $\bL$ is the bolometric
correction factor correlating this specific luminosity to the total gamma-ray
luminosity of the burst. The value of $\bL$ depends essentially on the high
energy cutoff in the photon emission, which cannot be inferred from current
data; for internal shocks one expects $\hat\eps \lsim 100\MeV$ in the
comoving frame, corresponding to a canonical value $\bL\approx 10$.  In the
internal shock model, constraints on the bulk Lorentz factor can be inferred
from the requirement that the dissipation radius is larger than the radius of
the photosphere, $\rph \sim [10^{18}\cm]\,\LGRBobs \eta^{-3}$, below which
the wind is optically thick, and the radius of the external termination
shock, $\rter \sim [10^{18}\cm]\,\LGRBobs^{1/3}\Tburstn^{1/3}\eta^{-2/3}$,
which requires the bulk Lorentz factor of the ejecta, $\Gamma\sim \eta$, to
be in the range $30\lsim \Gamma \lsim 10^3$, leading to magnetic fields in
the range $10^7\G \gsim B\gsim 1\G$, assuming equipartition between magnetic
field and photons, standard values for $\Lbobs$ and $\bL$, and typical time
scales of $0.1\scnd$ for short-term variability, and $30\scnd$ for the total
duration of a featureless burst. The latter values, $\Gamma\sim 10^3$ and
$B\sim 1\G$, would imply that the internal shocks occur on similar time
scales and physical conditions as the external shock, making both scenarios
virtually identical with respect to the discussion of transients. Typical
parameters for assumed internal shocks with $\cT\ll\Tburst$ are $\Gamma\sim
300$ and $B\gsim 10^3\G$, which are also required by models predicting the
acceleration of UHECR protons in this scenario \cite{Wax95,RM98hunt}.

We still need to relate the parameters $R$, $\xL$ and $\cD$, defined in
\RefSec{theory:Epmax} for a causally connected emission region moving at some
angle $\phiview$ to the line of sight, to the parameters describing the
expanding flow in a GRB.  In the comoving frame of the wind (at an arbitrary
point), the apparent thickness of the wind zone extending to a radius $r$ is
$r/\Gamma$. Similarly, the transversal extent of causally connected regions
in the comoving frame of the flow is $r/\Gamma$, because regions farther
apart move away from each other with velocities larger than $c$ ($r/\Gamma$
may thus be interpreted as the ``Hubble radius'' of the expanding emission
region, \cf\ \cite{MRees93apj}). If the energy dissipation takes place at a
radius $\rd$ where the bulk Lorentz factor is saturated, $\Gamma \sim \eta$,
the comoving ``linear size'' of the emission region can thus be written as $R
\sim \rd/\eta$. On the other hand, the isotropic luminosity of the burst in
the observer frame is related to the photon energy density in the comoving
frame of the wind by $\cL = 4\pi\rd^2 c \Gamma^2 \uph$, which is identical to
\Eq{theory:lum} in conjunction with the linear boosting formula, $\cL =
L\cD^4$, if we use the parameters $R=\rd/\Gamma$, $\xL=1$ and $\cD =
\Gamma$. For $\Gamma\sim\eta=\const$, all parameters are therefore correctly
derived if we treat the GRB as emission from a spherical region of radius
$R\sim\rd/\eta$ boosted with a Doppler factor $\cD\sim \eta$. Moreover, if
the gamma-ray emission originates from synchrotron radiation of electrons
whose cooling time scale is much shorter than the crossing time
\cite{ReesM94}, we can also apply $R \approx c \cT \cD$, or $\rhoT \approx
1$; since we can apply our discussion to both internal spikes in strongly
variable bursts and to featureless bursts in total, we allow the value of the
normalized time scale in the broad range $10^{-3}<\TGRBobs<100$. According
to our definition of the radius $R$ as the ``Hubble radius'' of the expanding
emission region, we obviously also have $\bad = 1$, and the effective space
fraction available for the gyration of protons is essentially constrained by
adiabatic losses, $\cgyr\sim (\pi\alpha\thF)^{-1}$, where both $\alpha=1$ and
$\alpha=2$ have to be considered possible values, depending on whether the
magnetic field is largely transversal or isotropized by reconnection
\cite{CThom94}.

\subsub{spec}{GRB neutrino spectrum and maximum energy}

The observed electromagnetic spectrum of a GRB can be approximately described
as a broken power law, with a break energy $\ebgrbobs \sim 300\keV$ in the
observer frame. The photon number spectrum is then given by $N(\eps)
\nearprop \eps^{-2}$ for $\eps > \ebgrb$ (thus $a=2$), and for $\eps <
\ebgrb$ it is $N(\eps) \nearprop \eps^{-2/3}$, with $\ebgrb = \ebgrbobs
\cD^{-1}$.  This yields a proton break Lorentz factor in the range $10^4\lsim
\gb\lsim 5\mal 10^5$, corresponding to the range of possible Doppler factors
given above. When we consider neutrinos of energy $\Enuobs \gsim 100\TeV$, we
require proton Lorentz factors of $\gamma_p \gsim 3\mal 10^6/\cD$. For $\cD\sim
100$, we therefore have $\gamma_p \gsim \gb$; for simplicity, we restrict our
considerations to the case $\gamma_p\gg\gb$, noting that this might be only
marginally correct for the lower energy bound of the VHE neutrino regime. In
a $\eps^{-2/3}$ low energy spectrum, we simply have $\Ngtot/\Ngb = 4 -
3(\gb/\gamma_p)^{1/3}$ for $\gamma_p>\gb$; for simplicity, we use $\Ngtot/\Ngb
\approx 3$ for all $\gamma_p$. Introducing normalized quantities also for the
Doppler factor, $\cD = 100\,\cD_2 $, the magnetic field, $B = [10^3\G]\,B_3$,
and the bolometric correction factor, $\bL = 10\bGRB$, the coordinates of the
star-point of the parameter space, where all proton cooling processes have
equal time scales at the maximum proton energy, are then
\eqns{GRB:starpoint}
\subeq{D}   \Dstar_2     \approx 	&\mrbox{4}{ 0.9} 		
	&\times\; \LGRBobs^{1/4} \TGRBobs^{-1/4} \alpha^{-1/4}\\
\subeq{B}   \Bstar_3     \approx 	&\mrbox{4}{4.5}
	&\times\; \LGRBobs^{-1/6}\TGRBobs^{-1/2}\thF^{1/3} \alpha^{5/6}\\ 
\subeq{eqBg}\eqBgstar  \approx	&\mrbox{4}{2\mal10^{-2}}
	&\times\; \LGRBobs^{1/6}\TGRBobs^{-1/2}\thF^{2/3}\alpha^{1/6}\bGRB^{-1}
\quad.
\text
In the following we express the magnetic field by its equipartition
parameter. Figure~\reffig{GRB:parspace} shows the GRB parameter space and the
separate regions of dominance of the various cooling processes limiting the
neutrino energy, and the corresponding neutrino spectral shapes for a set of
possible parameters, including both millisecond flares in internal shocks and
featureless GRBs; unlike in the AGN case, we do not use parameters scaled
relative to the star-point, and put more emphasis on secondary particle
cooling, which we consider separately for pions and muons. We see that in all
cases, the energy of neutrinos from both pion and muon decay is limited by
secondary particle cooling, where synchrotron cooling of pions and muons
plays the most important role. Only in a limited part of the parameter space
does an additional break due to adiabatic cooling of muons appear in the
spectrum, while adiabatic cooling is generally unimportant for pions. At the
star-point, the maximum neutrino energy from Gamma-Ray Bursts is therefore
considerably below $\Enuobsstar$, so that we omit its value here to avoid
confusion.
\ifsubmit\relax
\else
\newdimen\figurewidth
\ifrevtex\figurewidth=0.88\textwidth\else\figurewidth=0.7\textwidth\fi
\setfig{p}{GRB:parspace}
\centerline{\epsfxsize=\figurewidth\epsfbox[0 122 555 780]{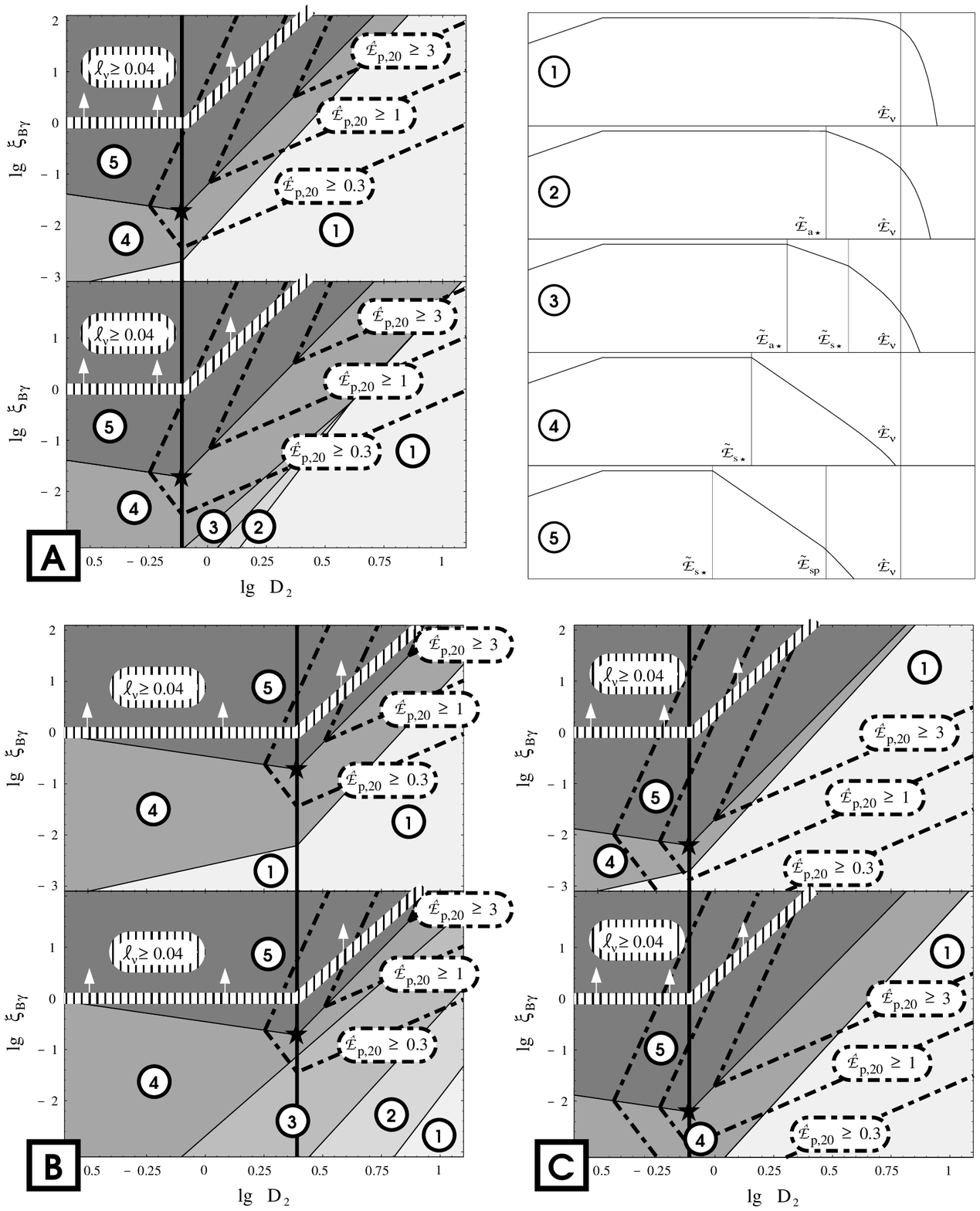}}
\figcap{\capsize Dominant cooling processes and spectral shapes for neutrino
production in Gamma-Ray Bursts --- {\bf Upper left and lower panel:}
Parameter space for three different parameter sets, all assuming
$\thF=\bGRB=1$, and $\alpha=2$: (A) canonical case, $\LGRBobs = \TGRBobs =
1$; (B) short intrinsic flares, $\cT=10\ms$, and $\LGRBobs=1$; (C) extreme
case for bright afterglow burst, $\cL=3\mal 10^{52}\erg\scnd^{-1}$ and
$\cT=30\scnd$ (which implies an isotropic bolometric photon energy
$\bL\Lbobs\cT = 10^{55}\erg$, requiring $\Omfrac\ll1$). The central line
divides regions where photohadronic cooling dominates adiabatic cooling (left
from the line), and vice versa (right from the line), the star-point of equal
proton cooling times at the maximum energy is indicated. Shaded regions
correspond to spectral shapes produced by the subsequent change of dominant
proton and secondary particle cooling processes: (1) adiabatic/photohadronic
cooling dominant up to $\Enumaxobs$; (2) $\Enumaxobssad<\Enumaxobs$; (3)
$\Enumaxobssad < \Enumaxobsssyn < \Enumaxobs$; (4)
$\Enumaxobsssyn<\Enumaxobs$; (5) $\Enumaxobsssyn<\Enubobspsyn<\Enumaxobs$. In
each figure, the upper part corresponds to neutrinos from pion decay and the
lower part for neutrinos from muon decay. Also indicated are the regions
allowing UHE cosmic ray production up to $\EpUHE$ (black chain lines), and
the region corresponding to the neutrino fluxes predicted in
Ref.~\cite{WB97}, for $\eqpB=1$ (white hatched line and arrows). {\bf Upper
right}: Schematic representation of corresponding spectral shapes, $\log
\LTnuobs(\Enuobs)$ vs. $\log \Enuobs$, corresponding to regions (1)--(5). The
lower break indicates the spectral change at about $100\TeV$ due to the
change of the photon target spectrum at $\gamma_p\sim \gb$ (see
Ref. \cite{WB97}), which is not discussed in this paper.}
\text\fi

Because of the dominance of secondary particle cooling, 
\Eqs{theory:Enumax:D:pimu} must be used to determine the neutrino energy
limit at a given Doppler factor, $\Enuobslim(\cD)$, which is reached along a
line $\eqBg = \eqBgssynstar (\cD/\Dstar)^5$ in the parameter space, with
\eqns{GRB:xiBssyn}
\subeq{pi} \eqBgpisynstar =& 2.3\mal 10^{-4} &\vphantom{1} \times 
	\LGRBobs^{1/4} \thF \TGRBobs^{-1/4} \alpha^{-1/4}\bGRB^{-1}\\
\subeq{mu} \eqBgmusynstar =& 1.7\mal 10^{-5} &\vphantom{1} \times 
	\LGRBobs^{1/4} \thF \TGRBobs^{-1/4} \alpha^{-1/4}\bGRB^{-1}\;.
\text
Recent observations of GRB afterglows at large redshifts require a total
isotropic energy emitted in photons of $\bL\Lbobs\Tburst \gsim 10^{53}\erg$
\cite{Kou97}. Since this value comes very close to the gravitational energy
released in the collapse of a compact object (\eg, a neutron star), it is
most likely that the energy of a GRB is not emitted isotropically. Moreover,
efficient hadronic emission of GRB has to assume that comparable amounts of
energy are present also in other channels, like magnetic fields or protons.
While collimation into jets is a distinct possibility, the evidence for it is
not as obvious as in AGN; in the following we assume that the GRB energy is
emitted into a ``firecone'' of solid angle $4\pi\gg\Omega\gg\cD^{-2}$, and we
introduce the normalized solid angle $\Omfrac\equiv\Omega/4\pi\ll 1$. For GRB
with a strongly variable light curve, the transients considered here are
single, isolated radiation spikes rather than the whole burst; if we assume,
however, that the total energy is equally distributed over the individual
flares and that there are no extended gaps in the light curve, we have
$\Etottrans/\bL\Lbobs\cT \sim \Eburst/\bL\Lbobs\Tburst$, which again allows a
common description of both subflares within GRB and featureless bursts. For
simplicity, we assume in the following parameters which allow relativistic
protons and magnetic fields to dominate the energy, $\eqpB\sim 1$ and
$\eqBgmax \sim \Omfrac^{-1} \LGRBobs^{-1} \gg 1$, and neglect the dependency
on the weakly varying factors $\alpha$ and $\bGRB$, leading to
\eqns{GRB:Dlim}
\subeq{pi} \Dlim_{2,\pi} &\approx& 
	5 \times \Omfrac^{-1/5} \TGRBobs^{-1/5} \thF^{-1/5}\\
\subeq{mu} \Dlim_{2,\mu} &\approx& 
	8 \times \Omfrac^{-1/5} \TGRBobs^{-1/5} \thF^{-1/5}\quad,
\text
which means that we only need to discuss the case $\Dlim_\mu>\Dstar$. We then
obtain for the neutrino energy limit
\eqns{GRB:limits:E}
\subeq{pi}  \Enuobspilim(\cD_1) \approx& \mrbox{12}{1\mal 10^{18}\eV
		\times D_2^{3/2} \TGRBobs^{1/2} \thF^{-1/2}} 
		& < 1\mal 10^{19}\eV\times
			\Omfrac^{-3/10} \TGRBobs^{1/5}\thF^{-4/5}\\
\subeq{mu}  \Enuobsmulim(\cD_1) \approx& \mrbox{12}{4\mal 10^{17}\eV
		\times D_2^{3/2} \TGRBobs^{1/2} \thF^{-1/2}} 
		& < 1\mal10^{19}\eV\times
			\Omfrac^{-3/10} \TGRBobs^{1/5}\thF^{-4/5}\quad.
\text
It therefore seems that photohadronic neutrinos from GRBs can reach energies
up to $10^{19}\eV$ and above. However, the corresponding normalized neutrino
luminosities at the neutrino energy limit are
\eqns{GRB:limits:LR}
\subeq{pi} \LRnulimpi(\cD_2) \approx& 2\mal 10^{-6} \times D_2 \thF 
		& < 1\mal 10^{-5} \times
			\Omfrac^{1/5} \TGRBobs^{-1/5}\thF^{4/5}\\
\subeq{mu} \LRnulimmu(\cD_2) \approx& 2\mal 10^{-7} \times D_2 \thF 
		& < 1\mal 10^{-6} \times
			\Omfrac^{1/5} \TGRBobs^{-1/5}\thF^{4/5}\quad,
\text
which shows that even in the limiting case $\cD=\Dlim$ only very small
fractions of the photon energy can be emitted in neutrinos at the limiting
energy. Since we assumed $\eqpB \sim 1$ and $\eqBgmax\gsim 1$ in our
calculations, the low values of $\LRnulim$ are clearly due to a very low
efficiency along the line $\eqBg = \eqBgssyn(\cD)$. Assuming a neutrino
conversion efficiency of $\Xnu\sim 0.2$, $\eqBg\sim 1$ and $\up \sim 5
\epsb^2 (d\Nph/d\eps)_{\rm b}$, which is consistent with our equipartition
assumption, $\eqpB\sim \eqBg \sim 1$, for $\bL\sim \bp\approx 5$, \viz,
$\LRnu\sim0.04$, Waxman and Bahcall \cite{WB97} derived a GRB-related
neutrino event rate at $\Enuobs \gsim 10^{14}\eV$ of about $10{-}100$ per
year in a $\km^3$ detector. This event rate would be below the expected
background, but still statistically significant because of the time and
directional correlation to the bursts.  It is clear from the above that if we
try to maximize the neutrino energy in the UHE range, following
\eq{GRB:limits:E}, this estimate would have to be reduced by more than 4
orders of magnitude, leading to insignificant event rates at these highest
energies even if the usually larger detector volume of UHE experiments is
taken into account.

We can also turn the question around and ask which is the region of parameter
space where neutrino fluxes of the magnitude predicted by Waxman and Bahcall
\cite{WB97} are expected, \ie, $\LRnu\gsim 0.04$, and then derive the
maximum neutrino energy for these parameters. The assumptions of Waxman and
Bahcall for a single burst (or sub-flare) can be essentially put in the form
$\LTnuobs\sim 0.2\Lbobs\cT$, which can be rewritten in the form of
\Eq{theory:LRnumax} as $\bL\Xnu\eqBg\eqpB\gsim 0.2\bp$. To optimize the
efficiency, we have to evaluate $\Xnu$ at an energy where it is not yet
diminished by synchrotron cooling of either protons or secondary particles;
since $\cD>\Dstar$, this means that we have to use \Eq{theory:Xnu:ad} for
$a=1$. For $\bp\sim \bL$ and $\eqpB\sim 1$ this leads to the condition
$\eqBg\gsim (\cD/\Dstar)^4$ for $\cD\ge\Dstar$; for $\cD<\Dstar$, the
efficiency is constant since photohadronic cooling
dominates. Figure~\reffig{GRB:parspace} shows that this region of the
parameter space is entirely enclosed by the regions of pion, muon and proton
synchrotron cooling dominance, which means that $\Enumaxobsssyn$ has to be
used as the maximum neutrino energy, according to our assumption of adiabatic
cooling dominance. Inserting into \Eq{theory:Enumax:pimu:syn} we obtain
\eqns{Enumax:WB97flux}
\Enumaxobspi <& 7\mal 10^{16}\eV \times \cD_2^2 \TGRBobs^{1/2}
	&< 6\mal10^{16}\eV \times \Omfrac^{-1/2}\\
\Enumaxobsmu <& 5\mal 10^{15}\eV \times \cD_2^2 \TGRBobs^{1/2}  
	&< 4\mal10^{15}\eV \times \Omfrac^{-1/2}\quad,
\text
where the second limit was derived using $\eqBg< \Omfrac^{-1}\LGRBobs^{-1}$
and $\Omfrac\ll 1$. This means that Gamma-Ray Bursts can still produce UHE
neutrinos with considerable fluxes, but probably not above
$10^{18}\eV$. 

While this paper was in preparation, it has been proposed by Vietri
\cite{Vie98prl} that neutrinos of more than $10^{19}\eV$ can be produced in
external shocks of GRB and in GRB afterglows, with fluxes observable in very
large scale air shower experiments. Concerning the external shock of the main
burst, this prediction is clearly in conflict with our upper limit stated in
\eqs{GRB:limits:LR} and \refeq{Enumax:WB97flux}. In the afterglow, the larger
time scales and lower photon energies make the situation rather comparable to
blazar jets, and \eqs{Enumax:WB97flux} do not apply; on the basis of the
considerations presented here, we cannot rule out the possibility of
producing neutrinos of such extreme energies in afterglows. It has been
recently shown that such large UHE neutrino fluxes would also be in conflict
with the assumption that (a) the cosmic ray production spectrum is $\propto
\gamma_p^{-2}$, as assumed also here, and (b) that the locally observed
cosmic ray energy density above $10^{19}\eV$ is homogeneous throughout the
universe and does not evolve with cosmological redshift \cite{WB98}.
 
\subsub{cr}{The relation between cosmic ray and neutrino maximum energies}

GRBs have been proposed as possible sources for the highest energy cosmic
rays \cite{Wax95,Vie95}, which are observed up to $\Epmaxobs\sim
3\mal10^{20}\eV$ \cite{Bie97jphG}. The maximum proton energy in the observer
frame is $\Epmaxobs = m_p c^2 \gpmax \cD$; at the star-point of the parameter
space, we find $\Epmaxobs^* \approx [1\mal
10^{20}\eV]\,\LGRBobs^{1/3}\thF^{-2/3}\alpha^{-2/3}$.  The highest proton
energy for a given Doppler factor is obviously achieved at the line of equal
time scales for proton synchrotron colling and adiabatic losses (the same
condition which determines the maximum neutrino energy if secondary particle
cooling plays no role). This is the border line of region (5) in the
parameter space shown in \fig{GRB:parspace}, so we find $\Epobslim =
\Epmaxobs^* (\cD/\Dstar)^{4/3}$ for $\cD\ge\Dstar$, leading to
\eqn{Epstar}
\Epobslim(\cD_2) \approx 1\mal10^{20}\eV\times \cD_2^{4/3} 
\TGRBobs^{1/3}\thF^{-2/3}\alpha^{-1/3}\quad. 
\text
Since $\eqBg\propto \cD^{14/3}$ along this line, the required magnetic field
equipartition parameter rises fast; as a function of the proton energy in the
observers frame, $\EpUHE$, in units of $10^{20}\eV$, we can formulate the
minimum requirements in $\cD$ and $\eqBg$ as 
\eqns{cond:UHECR}
\DminUHE(\EpUHE) &\approx& \mrbox{3}{90} \times 
	\EpUHE^{3/4}\TGRBobs^{-1/4}\thF^{1/2}\alpha^{1/4}\\
\eqBgUHE(\EpUHE) &\approx& \mrbox{3}{0.01} \times 
	\EpUHE^{7/2}\TGRBobs^{-1/2}\thF^{3}\alpha^{5/2}\LGRBobs^{-1}\bGRB^{-1} 
\quad.
\text
In correspondence with the result of Waxman \cite{Wax95}, we find that bulk
Lorentz factors $\cD\gsim 300$ and magnetic fields close to equipartition
with the radiation ($\eqBg\sim 1$) are sufficient to produce the highest
energy cosmic rays, $\EpUHE \gsim3$, provided that protons are accelerated
on their Larmor time scale. Assuming minimal conditions for the production of
UHE cosmic rays, \ie, $\cD=\DminUHE(\EpUHE)$ and $\eqBg=\eqBgUHE(\EpUHE)$, we
can derive the corresponding maximum neutrino energy as a function of
$\EpUHE$ as
\eqns{Enumax:Epmax}
\subeq{pi} \Enumaxobspi\Big|_{\Epmaxobs = \Epobslim(\cD)} 
	\sim& 5\mal 10^{17}\eV & \vphantom{1} \times 
	\EpUHE^{5/4}\thF^{1/2}\TGRBobs^{1/4}\alpha^{-1/4}\\
\subeq{mu} \Enumaxobsmu\Big|_{\Epmaxobs = \Epobslim(\cD)} 
	\sim& 3\mal 10^{16}\eV & \vphantom{1} \times 
	\EpUHE^{5/4}\thF^{1/2}\TGRBobs^{1/4}\alpha^{-1/4}
\text
where we again assume $\Enumaxobss = \Enumaxobsssyn$. We see that, although
secondary particle cooling limits the neutrino energies to values much below
the canonical $\Enumaxobs = 0.05\Epmaxobs$ assumption, the maximum neutrino
energy still has a tendency to increase with $\Epmaxobs$. However, 
\eqs{Enumax:Epmax} assume just minimal conditions for the production of UHE
cosmic rays, while the region of the parameter space corresponding to
$\Epmaxobs\gsim [10^{20}\eV]\,\EpUHE$, as shown in \fig{GRB:parspace}, allows
neutrino break energies principally somewhat above or below the value stated
in \eqs{Enumax:Epmax} for given $\EpUHE$.

We note, however, that according to the standard assumptions of shock
acceleration, neutrino production and cosmic ray ejection from magnetically
confined sources are physically connected processes. Although it is a
distinct possibility that cosmic ray acceleration to $\Epmaxobs>10^{20}\eV$
on the one hand, and efficient VHE neutrino production requiring only cosmic
rays of lower energy on the other, may happen at different radii in the
expanding shell \cite{WB98}, the {\em ejection} of UHE cosmic rays is
non-trivial in this scenario, since the particles are advected downstream,
and thus accumulate inside the expanding shell and remain magnetically
confined. As the shell continues to expand and the comoving magnetic field
decreases, the cosmic rays would lose most of their energy by adiabatic
cooling, before they eventually escape when the shock slows down. The easiest
way to circumvent this problem, and to eject cosmic rays with the high
energies they receive at the shock, is to convert them into neutrons in, \eg,
$p\gamma\to n \pi^+$ reactions, which decouples them instantaneously from the
shell as long as their reconversion probability is low \cite{RM98hunt}, which
is the case for $\cD>\Dstar$ (\cf\ \RefSec{phothad:pcool}). This means that a
low neutrino production efficiency corresponds to a low cosmic ray ejection
efficiency for the standard scenario of shock acceleration, assuming magnetic
confinement of all accelerated charged particles. This physical connection of
both processes makes the hypothesis that Gamma-Ray Bursts are the sources of
the highest energy cosmic rays testable by neutrino VHE and UHE observations,
both with respect to the flux and the maximum energy of the putative GRB
correlated neutrino spectrum. It is also obvious from \fig{GRB:parspace},
that this joint scenario, \ie, $\EpUHE\ge 3$ and $\LRnu\ge 0.04$, also
requires extreme values of the GRB parameters $\cD$ and $\eqBg$.

\sect{concl}{Conclusions}

We have presented a detailed investigation of the production processes of very
energetic (${\gsim} 10^{14}$ eV) photohadronic neutrinos in relativistically
boosted astrophysical sources. Using the constraints set by the source
variability, and assuming that the acceleration process for protons is of the
Fermi type, we derived limits on the maximum energy and the position of
possible breaks in the neutrino spectrum. Comparing the effects of various
cooling processes for both protons and secondary particles in the hadronic
cascade leading to neutrino production (\ie, pions and muons), we find a
general upper limit on the neutrino maximum energy, which does only depend on
the Doppler factor of the emission region relative to the observer. Energetic
constraints allow one to turn this into a general upper limit, which is only
dependent on observational parameters of the transient, but not on any model
dependent parameters. This is the major result of this paper. In some cases,
and assuming that the energy in protons, magnetic field and photons are near
equipartition, stricter limits can be imposed when considering both neutrino
energy and expected flux. We apply this general result to two classes of
proposed cosmic neutrino sources: hadronic models of Doppler beamed jets from 
Active Galactic Nuclei (AGN), also called blazars, which are known to emit 
most of their energetic radiation in short, distinct flares, and Gamma-Ray 
Bursts (GRBs).

For blazar flares, we confirm that under the most common assumptions the
neutrino energy is limited to ${\lsim} 10^{18}\eV$ by Larmor radius (or
adiabatic) constraints of the accelerated protons. For short (${\lsim} 1\hrs$)
flares, however, the maximum energy of neutrinos from muon decay may be
additionally suppressed by muon synchrotron cooling. Assuming that the AGN is
fueled by Eddington limited accretion on a supermassive Black Hole with
$M_{\rm\SSS BH}\le10^{10}\Msol$, we show that neutrinos from AGN flares
cannot exceed energies of ${\sim}10^{19}\eV$, independently of the time scale
of the flare. For electron neutrinos, which result only from muon decay, we
can show that fluxes of the magnitude usually assumed in the literature (\ie,
similar to the X-ray luminosity of the source) can only be attained if the
energy extends only up to about $10^{18}\eV$. Unless vacuum neutrino
oscillations occur in nature, this has important implications for the
neutrino event rates expected in the Pierre Auger Observatory, where electron
neutrinos are proposed to be most easily detected because of the distinct
properties of their induced air showers.

For GRBs we find that the synchrotron cooling of pions
and muons limits the maximum neutrino energy over most of the allowed region
of parameter space. We show that, although neutrinos from GRBs are in
principle able to exceed $10^{19}\eV$, in particular for acceleration over
long time scales in or near external shocks, this possibility would imply
extremely low efficiencies and thus insignificant neutrino fluxes.  If we
require that neutrinos are also produced with fluxes similar to the
gamma-ray flux, and applying the usual energetic constraints for
near-isotropic GRB sources, we find an upper limit on the neutrino energy of
${\lsim}10^{17}\eV$ for muon neutrinos (from pion decay), and ${\lsim}
10^{16}\eV$ for electron neutrinos. This limit can only be increased to UHE
(${>}10^{17}\eV$) neutrino energies if strongly collimated outflows are
assumed. We also show that the conditions for GRBs to accelerate protons up
to the highest energies observed in cosmic ray air shower experiments (${\sim}
3\mal10^{20}\eV$) coincide with the conditions for efficient neutrino
production, and expect a flat neutrino power spectrum extending up to a break
energy in the range of $10^{16}{-}10^{18}\eV$. If neutrino production and
cosmic ray ejection from GRB are connected processes, as implied by the
standard assumption of magnetic confinement of shock accelerated particles in
an expanding shell, this would make the hypothesis that Gamma-Ray Bursts are
the sources of the observed ultra high energy cosmic ray spectrum testable
with neutrino observatories.

As a corollary of our investigation of the relevance of secondary particle 
cooling to hadronic cascades in common models of astrophysical transients, we 
have also checked the effect of this process on other proposed, non-transient 
sources for cosmic neutrinos. In particular the predicted diffuse neutrino 
background from AGN cores, frequently used for event estimates in high energy 
neutrino detectors, was previously derived disregarding secondary particle 
cooling. Here we obtain for energies above $10^{15}\eV$ a strongly reduced 
contribution and a lower cutoff in the electron neutrino component, and 
a reduction of about a factor $3$ for the expected diffuse muon neutrino flux 
(see \RefApp{AGNcore}).

In cosmic sources where the neutrino energy is limited by secondary particle
cooling, which is clearly predicted for Gamma-Ray Bursts and is possible in
some blazar flares, the expected difference in the cutoff energy of electron
and muon neutrinos could also serve as a laboratory to test the existence of
neutrino vacuum oscillations in nature at very high energies --- a detected
change in the neutrino composition near the cutoff energy could rule out this
possibility. Such a measurement would require a large detector sensitive in
the range $10^{15}{-}10^{18}\eV$, capable of detecting both electron and muon
neutrinos and able to distinguish between flavors.

\ifrevtex\acknowledgements 
\else\section*{Acknowledgements}\fi%
We wish to thank K.~Mannheim and E.~Zas for discussions. We also thank
P.L.~Biermann, T.~En{\ss}lin, T.K.~Gaisser, and the anonymous referee for the
careful reading of the manuscript, and helpful comments and suggestions. This
work was supported in part by NASA grant NAG5-2857.

\begin{appendix}

\sect{APPENDIX:notation}{Notational conventions}

Below we explain some general conventions we use throughout the
paper. Table~\reftab{symbols} lists some generally used symbols; it does not
contain symbols which are used only in the section where they are defined.

\ifsubmit\relax
\else
\newpage
\settab{h}{symbols}
\tabcap{Index of frequently used symbols.}
\begin{tabular}{lp{0.46\textwidth}@{\quad}p{0.24\textwidth}l}
symbol 		& meaning	& definition/relations 	& introduced in\\\hline
$\mx$, $\rx$, $\gammax$ 
	& particle mass, classical radius, and Lorentz factor \rule{0pt}{2ex} 
	[$\SS\bullet{=}p,\pi,\mu$] 
	& $\rx = e^2/\mx c^2$
	& general\\ 
$\taux$, $\tauxRF$  
	& unstable particle lifetime [$\SS\bullet{=}n,\pi,\mu$] 
	& $\taux = \gamma_\bullet \tauxRF$ & \RefSec{phothad:pcool} \\  
$\gsprod$, $\gsdec$ 
	& Lorentz factor of secondary particles at production,
	decay\tablenotemark[2] 
	& $\gmudec{\le}\gmuprod{\approx}\gpidec{\le}\gpiprod{\approx}\gamma_p$
	& \RefSec{phothad:pcool} \\
$\Ngtot$, $\Ngb$, $\epsb$, $a$ 
	& total photon density, density and power law index above $\epsb$
	& $d\Ngtot = \eps^{-a}\,d\eps$, $\eps>\epsb$  
	& \RefSec{phothad:pcool}\\
$\eth$ 
	& threshold photon energy in proton RF for $\pi$-production  
	& \Eq{APPENDIX:phothad:Etapi}
	& \RefSec{phothad:pcool}\\
$\gb$, $\gbt$ 
	& characteristic Lorentz factor for power law approximation 
	& $\gb = \eth/2\epsb$, $\gbt = \gb/\cD$
	& \RefSec{phothad:pcool}\\
$\tpib$, $\Etapi$
	& pion production cooling time for $\gamma_p=\gb$, effective
	inelasticity weighted cross section for power law photon spectrum
	& $\tpib = [c\Etapi\Ngb]^{-1}$,\newline
	\Eqs{phothad:tpi}, \RefEq{theory:tpib:lum}, 
	\RefEq{APPENDIX:phothad:Etapi} 
	& \RefSec{phothad:pcool}\\ 
$B$, $\omBx$ 
	& magnetic field, particle cyclotron frequency 
	& $\omBx = eB/\mx c$ 
	& \RefSec{phothad:pcool}\\
$\tsynx$, $\tad$ 
	& particle synchrotron [$\SS\bullet{=}\pi,\mu,p$] and adiabatic cooling
	time\tablenotemark[1] 
	& \Eq{phothad:tsyn}, $\tad = 2|B/\dot B|$ 
	& \RefSec{phothad:pcool}\\ 
$\tpg$, $\tpi$, $\tescn$, $\tBHp$ 
	& total and specific photohadronic cooling times\tablenotemark[1]
	& $\tpg^{-1} = \tpi^{-1}+\tescn^{-1}+\tBHp^{-1}$
	& \RefSec{phothad:pcool} \\
$\tpcool$ 
	& total proton cooling time scale 
	& $\tpcool^{-1} = \tad^{-1}+\tsynp^{-1}+\tpg^{-1}$ 
	& \RefSec{phothad:pcool} \\
$\fpg$, $\fsyn$, $\fad$, $\fmax$ 
	& rate of proton cooling relative to pion production 
	& \Eq{phothad:tcool/tpi} 
	& \RefSec{phothad:pcool} \\
$\Xnu$, $\Xnus$    
	& efficiency for neutrino production, from specific
	decay\tablenotemark[2] 
	& \Eq{phothad:Xnu} 
	& \RefSec{phothad:pcool} \\
$\Upint$, $\bp$, 
	& total {\em injected} proton energy density, bolometric correction
	factor 
	& \Eq{phothad:Up}
	& \RefSec{phothad:specshape}\\
$\gpmax$, $s$ 
	& maximum proton Lorentz factor, power law index
	& $d\dot N_p\propto\gamma_p^{-s}\,d\gamma_p$, 
	$\gamma_p\lsim\gpmax$
	& \RefSec{phothad:specshape} \\
$\bar L_\nu(E_\nu)$, $\Enumax$, $q$
	& neutrino emission spectrum, cutoff energy, local spectral index 
	& \Eqs{phothad:specnu}
	& \RefSec{phothad:specshape} \\
$\Gamma$, $\bflow$, $\phiview$, $\cD$ 
	& bulk Lorentz factor, velocity in units of $c$, viewing angle and
	Doppler factor of the emission region in observers frame 
	& $\bflow = \sqrt{1-\Gamma^{-2}}$,\newline 
		$\cD = [\Gamma(1-\bflow\cos\phiview)]^{-1}$ 
	& \RefSec{theory}\\ 
$\cT$, $\Lb$, $\bL$ 
	& observed duration of transient, luminosity at $\veps=\epsbobs$ 
	& $T = \cT\cD$, $\Lbobs = \Lb\cD^4$, \Eq{theory:tics}
	& \RefSec{theory}  \\ 
$\Tcr$, $\Tinj$, $\Trad$ 
	& transient crossing, proton injection and radiative time scale 
	& \Eq{theory:Ttot}
	& \RefSec{theory:times:causality}\\ 
$R$, $\Rpar$, $\Rperp$ 
	& linear size of transient emitter, ${\SS\|}=$ in line of sight, 
	$\SS\perp=$ projected 
	& $\Rpar = c\Tcr$ 
	& \RefSec{theory} \\
$\rhoT$, $\xL$
	& geometric correction factors
	& \Eq{theory:Rlim}, \RefEq{theory:lum}
	& \RefSec{theory:times:causality}\\
$\rgyr$, $\tgyr$ 
	& proton Larmor radius and time 
	& $\rgyr = E_p/eB$, $\tgyr = 2\pi\rgyr/c$ 
	& \RefSec{theory:Epmax}\\
$\tacc$, $\thF$ 
	& acceleration time scale, normalized to Larmor time 
	& $\tacc = \thF \tgyr$ 
	& \RefSec{theory:Epmax}\\
$\alpha$, $\bex$, 
	& magnetic field decay parameter, expansion velocity of transient 
	& $B\propto R^{-\alpha}$, $\bex = \dot R/c$ 
	& \RefSec{theory:Epmax:larmor} \\
$\Enumaxobsgyr$, $\rhoL$ 
	& Larmor limit for neutrino energy, correction factor 
	& \Eq{theory:Enumax:Larmor},
	$\rhoL{\lsim}\min(\frac13,\frac1{\pi\thF\alpha\bex})$  
	& \RefSec{theory:Epmax:larmor}\\
$\Enumaxobspsyn$, $\Enumaxobspg$
	& neutrino cutoff energy limited by proton cooling\tablenotemark[1]
	& \Eqs{theory:Enumax:psyn}, \RefEq{theory:Enumax:pg} 	
	& \RefSec{theory:Epmax}\\
$\omdecs$ 
	& characteristic frequency for secondary particle decay\tablenotemark[2]
	& $\omdecs = \frac32 \sqrt{c/\tausRF\rs}$ 
	& \RefSec{theory:pimu} \\
$\Enumaxobssad$, $\Enumaxobsssyn$
	& critical neutrino energies for secondary particle
	cooling\tablenotemark[1]\tablenotemark[2] 
	& \Eqs{theory:Enumax:pimu:ad}, \RefEq{theory:Enumax:pimu:syn}
	& \RefSec{theory:pimu} \\
$\Enudecs$
	& neutrino energy in decay frame\tablenotemark[2]
	& $\Enus = \gs\Enudecs$ 
	& \RefSec{theory:pimu} \\
$\UpsT$, $\XiT$ 
	& dimensionless characteristic parameters of transient 
	& \Eqs{theory:UpsT}, \RefEq{theory:XiT}
	& \RefSec{theory}\\
$\uB$, $\ue$, $\uph$ 
	& magnetic, electron and photon energy density in emission region
	& $\uB = B^2/8\pi$, $\ue\sim\uph$
	& \RefSec{theory:modind:parspace} \\
$\eqBg$, $\eqpB$
	& energy equipartition parameters
	& \Eq{theory:eqBg}, $\eqpB=\Upint/\uB$
	& \RefSec{theory:modind:parspace} \\
$\Dstar$, $\Bstar$, $\gpstar$
	& ``star-point'' parameters\tablenotemark[3], all specific proton
	cooling times ${=}\tacc$  
	& \Eqs{theory:starpoint}
	& \RefSec{theory:modind:parspace} \\
$\Enuobsxlim(\cD)$, $\Bxsyn$
	& neutrino energy limit and corresponding $B$ for given $\cD$
	[$\SS\bullet{=}p,\decsym$]\tablenotemark[2]
	& \Eqs{theory:Bpsyn}--\RefEq{theory:Enumax:D:pimu}
	& \RefSec{theory:modind:uplim} \\
$\LRnulimx$
	& relative neutrino luminosity for $\Enumaxobs=\Enuobsxlim$
	& \Eqs{theory:LRnumax:psyn}, \RefEq{theory:LRnumax:ssyn}
	& \RefSec{theory:modind:efficiencies}\\
$\Etottrans$, $\eqBgmax$
	& total energy budget of transient, corresponding maximum $\eqBg$
	& \Eq{theory:eqBgmax}
	& \RefSec{theory:modind:efficiencies} \\
\end{tabular}
\tablenotetext[1]{Cooling processes are: adiabatic losses ($\rm\SS ad$),
synchrotron radiation ($\rm\SS syn$), photohadronic interactions ($\SS
p\gamma$), (charged) pion production ($\SS \pi,\pi^\pm$), Bethe-Heitler
$e^\pm$ pair production ($\rm\SS BH$), neutron escape ($\rm\SS esc,n$).} 
\tablenotetext[2]{The subscript ``$\SS\decsym$'' denotes secondary particles in
hadronic cascade here and throughout the paper  [$\SS\decsym=\pi^\pm,\mu$]}
\tablenotetext[3]{The superscript ``*'' generally denotes quantities taken at
the star-point of the parameter space, \ie, $\cD=\Dstar$, $B=\Bstar$, and
$\gamma_p = \gpstar$.}
\text
\fi

\paragraph{Units and normalized quantities:} We use cgs units, except for
particle energies which are given in standard multiples of electronvolts
($\!\eV,\MeV,\GeV,\TeV$), and particle interaction cross sections measured in
microbarn ($\mubarn$). In numerical calculations we use quantities normalized
to common powers of their standard unit (${\rm stu}$), $X_n \equiv
X/10^n\,{\rm stu}$ (\eg, $L_{51} = L/10^{51}\erg\scnd^{-1}$). Dimensionless
quantities may be normalized in common powers as well. This convention is
used consequently in \RefSec{astro}, which means that numerical subscripts
always denote normalization powers.

\paragraph{Reference frames:} Three reference frames are used in the paper:
the observers frame, the comoving frame of the emission region (relativistic
flow), and the rest frame of an interacting massive particle. Quantities are by
default given in the comoving frame; quantities in the observers frame are
denoted by calligraphic letters (\eg, $\cE$, $\cT$, $\cL$, $\eps$) of the same
kind as corresponding quantities in the comoving frame ($E$, $T$, $L$,
$\veps$). Quantities in the particle rest frame are denoted with a
superscript {\sc RF}.

\paragraph{Luminosity convention:} Luminosities quoted in the paper always
mean the isotropic radiated power at specific energy (frequency),
\eg,, $L = \eps^2(d\Nph/d\eps)$ or $L_\nu = \Enu^2 (d\Nnu/d\Enu)$. If we refer
to bolometric luminosities, we do this by explicitly multiplying with a
bolometric correction factor, \eg $L_{\rm bol} \equiv \bL\Lb$ (see
\tab{symbols}). The energy output of a transient over its time scale $T$ is
denoted the {\em time integrated luminosity}, $\bar L$.

\sect{APPENDIX:phothad}{Photohadronic interactions}

The major photohadronic interaction channels of protons are single pion
production with and without isospin flip, $p\gamma\to n\pip$ and $p\gamma\to
p\piO$, followed by several two-pion production channels, and multi-pion
production which dominates at very high interaction energies. Secondary
neutrons can contribute negative pions from $n\gamma\to p\pim$ reactions,
which are otherwise only produced in two-pion and multi-pion channels.  The
subsequent decay of the pions leads to neutrino production by
\eqns{decay}
\subeq{pi} \pipm &\to& \mupm\nu_\mu(\bar\nu_\mu)\\
\subeq{mu} \mupm &\to& \epm\bar\nu_\mu\nu_e(\nu_\mu\bar\nu_e)\quad.
\text
The charge of the initial pion is only relevant for the $\nu:\bar\nu$ ratio;
it plays a role for the electron neutrino component at energies $\Enuobs
\approx 6\mal10^{15}\eV$, where the detectability of $\bar\nu_e$ is enhanced
due to the $W^-$ resonance in interaction with atmospheric
electrons. Otherwise, neither underwater/ice, nor air shower experiments can
distinguish between neutrinos and anti-neutrinos, so that we can disregard
the sign of the pion charge. The average energy of the neutrinos is
determined by decay kinematics: it can be written as $\exval{E_\nu} =
\Enudecs\gs$, where $\Enudecs$ is the energy of the neutrino in the rest
frame of the decaying particle moving with Lorentz factor $\gs$. For pion
decay, $\Enudecpi = 30\MeV$ \cite{PDG96}, while in the three particle decay
of the muon $\exval{\smash{\Enudecmu}} \approx \frac13m_\mu c^2 = 35\MeV$; as
an approximation, we may use $\exval{\smash{\Enudecs}} \approx \frac14 m_\pi
c^2$.

At low interaction energies, $\ePRF\sim 340\MeV$, the photohadronic cross
section for both charged and neutral pion production is dominated by the
$\mD(1232)$ resonance, leading to single pion production with a $\pip:\piO$
ratio of $1:2$ following isospin symmetry, and $E_\pi\approx 0.2 E_p$ from
two-particle decay kinematics. These relations are often used as
characteristic for pion production, and we call it the $\mD$-approximation
\cite{GHS95,Ste73/79}; it is quite accurate for $\piO$ production, in
particular in steep photon spectra, and in thermal spectra with temperatures
$kT$ for proton Lorentz factors $\gamma_p \sim [340\MeV]/kT$, where higher
resonances and other photohadronic interaction channels have little or no
influence on the cross section. For charged pion production, however, other
processes contribute significantly both above ($\ePRF\gsim400\MeV$) and below
($\eth<\ePRF\lsim 250\MeV$) the dominance region of the $\mD(1232)$
resonance, and thus enhance the contribution of charged pions relative to the
$\mD$-approximation. These are in particular $N^*$ resonances at energies
above the $\mD(1232)$, but cross section data also require a contribution
from non-resonant pion production, which give an almost constant background
of about $100\mybarn$, extending from the immediate threshold (even before the
$\mD(1232)$ resonance becomes relevant) up to the highest energies
\cite{Laget81,Rac96PhD}. In the neutrino sources considered in this paper, a
proton spectrum extending up to a maximum energy $\Epmax = m_p c^2 \gpmax$,
interacts with an isotropic power law photon distribution, $d\Nph \propto
\eps^{-a}\,d\eps$, $a>1$, extending from a break energy $\epsb$ to some
cutoff at $\hat\eps\gg\epsb$, with a total number density $\Ngb$ of photons
with $\eps>\epsb$.  Below the break energy, we assume that the photon number
spectrum is flatter than $\eps^{-1}$ everywhere. We can then define the
inelasticity weighted effective cross section for pion production,
\eqn{Etapi}
\Etapi = 2\;\frac{a-1}{a+1}\int_1^\infty dx\,x^{-a}\;
	 \sum_{\rm i} \Big[\kappa_i \sigma_i\Big]_{\eps = x\eth}^{\rm\SSS RF}
	\quad\for a>1\;,
\text
where $\eth\approx 145\MeV$ is the threshold photon energy for pion
production in the proton rest frame, $\sigma_i$ is the cross section, and
$\kappa_i = \exval{\Delta E_p/E_p}_i$ is the proton inelasticity of the
reaction channel, evaluated at a photon energy $\ePRF=x\eth$ in the proton RF
(see \cite{Rac96PhD} for details). Introducing the characteristic Lorentz
factor $\gb = \eth/2\epsb$ of the protons, which is necessary to boost
background photons at the break energy above the reaction threshold for pion
production, we can then write the cooling time of protons with $\gamma_p =
\gb$ as $\tpib = [c \Ngb \Etapi]^{-1}$, where $\Ngb$ is the number
density of photons with $\eps>\epsb$. As a function of $\gamma_p$ the cooling
time $\tpi$ is then expressed by \Eqs{phothad:tpi}, and we note that all the
interaction physics, including the relative contributions of different
resonances or other reaction channels, are absorbed in $\Etapi$ and thus
independent of the proton energy --- this would not be the case in, \eg,
thermal photon spectra, where our results are not applicable. As a numerical
simplification, we also disregard the upper cutoff in the photon spectrum,
which is justified if the spectrum for $\eps<\epsb$ is sufficiently steep ---
for spectra with $a<2$, this approach is valid only in a limited range of
Lorentz factors below $\gb$ --- and use $\Etapi \approx \Etanorm/(a-1)$ for
$1.5<a\lsim 3$, with $\Etanorm \approx 22\mubarn$ \cite{Rac96PhD}.  If
$\gamma_p\gg\gb$, photohadronic interactions majorly happen at proton rest
frame energies far above the threshold, where the cross section and
efficiency is approximately constant; this justifies the approximation $\tpi
\approx [c\Ngtot\Etanorm]^{-1}$, used throughout the paper.

If no other cooling processes apply, the average number of pions produced per
proton can be found as $\exval{\Npi/\Np} = \Pitot/\Etapi \approx 7$, for
$1.5<a\le 3$, where $\Pitot$ is defined as in \eq{Etapi} by replacing the
proton inelasticity, $\kappa_i$, by the pion multiplicity of the reaction
channel. The average energy carried by each pion is then, independent of
$\gamma_p$, given by $\Epiav\!/E_p \approx \frac17 \approx m_\pi/m_p$, \ie,
the Lorentz factor can be treated as conserved in the interaction,
$\gpiprod\approx\gamma_p$. Here, $\gpiprod$ is the pion Lorentz factor {\em
at production}, which has to be distinguished from $\gpidec$, the pion
Lorentz factor {\em at decay}; for both pions and muons we consider the
possibility that they lose energy during their lifetime, \viz,
$\gsdec\le\gsprod$.  Because of the small mass difference between pion and
muon, we can also approximate $\gmuprod \approx \gpidec$. Distinguishing
between the charged pion and neutral pion multiplicity in the definition of
$\Pitot$ yields the charged pion fraction, $\pipm:\piO \approx 2:1$, almost
independent of the power law index $a$ \cite{Rac96PhD}. This result includes
all reaction channels, and is in contrast to the often used ratio $\pipm:\piO
= 1:2$, which is derived from the $\mD$-approximation. The discrepancy of a
factor of $4$ emphasizes the importance of charged pion production away from
the $\mD$-resonance in power law target photon spectra, which is also
relevant for the kinematics: it explains the difference of the usual
assumption, $\Epiav\approx \frac15 E_p$, (which is strictly valid for $\piO$
production, see above) and our result for charged pion production,
$\Epiav\approx \frac17 E_p$.

\sect{APPENDIX:nu:event}{Sensitivity range and energy dependence of neutrino
detector techniques}

The low detection probability for neutrinos above the TeV range requires
large detector volumes, which can be achieved, \eg, by the extension of
classical water Cherenkov detectors to larger dimensions, such as the NESTOR
or Lake Baikal (and the recently cancelled, pioneering DUMAND) experiments,
and similar projects in the planning stage \cite{GHS95}. The same technique
is also efficient in the deep antarctic ice, as shown impressively by the
recent detection of the first neutrino events by the AMANDA experiment
\cite{Hal98}, and there is hope to extend this detector to an effective
volume of $1\km^3$ in the future \cite{Hal97}. A cost-efficient way to
further increase the detector volumes could be the detection of radio pulses
\cite{FRMcK95} or acoustic waves \cite{BKM+96} from neutrinos in water or
ice, which however are limited to very high energies to obtain a reasonable
signal-to-noise ratio \cite{GHS95}.

The neutrino event rate per logarithmic energy interval in a given detector,
$\Fdet$, can be written as the product of the neutrino {\em number} flux,
$\LTnuobs/\Enuobs$, times the detection probability, $\Pdet$.  For deep
underwater/ice Cherenkov detectors, which are most efficient to detect muons
from $\nu_\mu\to\mu$ conversions because of the large mean free path of muons
in matter, the detection probability is essentially proportional to the ratio
of the muon mean free path to the neutrino mean free path, yielding
$\Pdet\nearprop \Enuobs^{0.8}$ for $\Enuobs\gsim10^{12}\eV$. At energies
above ${\sim} 10^{14}\eV$ the effective solid angle covered by the experiment
is reduced by earth shadowing effects \cite{NT96}, so that this technique
becomes ineffective for ultra high energies; it is also less sensitive to
electron neutrinos, because the short range of the electron in matter reduces
the effective volume. Radio Cherenkov detectors are proposed to work best in
deep ice. The detection probability is usually expressed as the effective
volume of the detector, increasing as $\Pdet\nearprop\Enuobs^{1.5}$ for
$\Enuobs\lsim 10^{16}\eV$, and is roughly constant above $10^{16}\eV$
\cite{FRMcK95}. The energy threshold for this technique is set by
signal-to-noise constraints at ${\sim} 5\mal10^{15}\eV$ \cite{GHS95}, thus in
the relevant sensitivity range the detection probability can be treated as
approximately constant. No clear predictions exist about the efficiency of
the acoustic method yet, which is probably most interesting at ultra high
energies.

Cosmic neutrinos may also cause air showers similar to cosmic rays, but
deeper penetrating and thus distinguishable due to their large zenith angle
\cite{ZHV93}. At ${\sim} 10^{15}\eV$, such ``horizontal'' air showers are
dominantly caused by atmospheric muons rather than cosmic neutrinos. Above
$10^{17}\eV$, however, the atmospheric background becomes low, and the air
scintillation technique used, \eg, in the HiRes Fly's Eye detector or the
Telescope Array \cite{ICRC97-UHECR}, largely improves the detectability
of horizontal air showers, providing much larger detection volumes than
underground detectors. Above ${\sim} 10^{19}\eV$, the planned Pierre Auger
Observatory \cite{Bor96} is expected to achieve considerable event rates,
using the same technique in conjunction with ground arrays for particle
detection \cite{PZ96}. The main caveat of the technique is the large
atmospheric background --- horizontal air showers produced by muon neutrinos
can be easily confused with air showers from atmospheric muons generated by
the prompt decay of charmed particles \cite{GHS95,ZHV93}. Horizontal air
showers produced by electron neutrinos have the unique property to be mixed
hadronic and electromagnetic showers \cite{ZHV93}, which allows to determine
distinctive triggering criteria for hybrid detectors like the Pierre Auger
Observatory, reducing the background \cite{PZ96}. This makes electron
neutrinos most interesting for UHE neutrino astronomy.

The detection probability for deeply penetrating horizontal air showers can
be expressed as the product of the hadronic neutrino cross section,
$\sigma_{\nu p}$, and the detector acceptance for an horizontal air
shower. Models predict correspondingly that $\sigma_{\nu p}\nearprop
\Enuobs^{0.5}$ for $\Enuobs \gsim 10^{15}\eV$ \cite{GQRS96}. The detector
acceptance for the Pierre Auger Observatory has been calculated as
$\nearprop\Enuobs^{0.3}$ for $\Enuobs\gsim 10^{17}\eV$, where horizontal air
shower detection is expected to be more efficient than other techniques
\cite{PZ96}. This gives rise to $\Pdet\nearprop\Enuobs^{0.8}$, which
is the same dependence as in the case in water/ice Cherenkov experiments.
For other air shower experiments, the dependence of the shower acceptance on
energy might be different, but a slow rise in the UHE regime seems to be a
common feature.

The expressions for the energy dependence of the detection probability are
highly approximate; exact results require expensive Monte-Carlo simulations,
considering the detailed properties of the experiment. However, we note that
the neutrino event rate per logarithmic energy interval, evaluated for the
most common detector techniques, follows closely the neutrino power spectrum,
$\Fdet/\LTnuobs \sim \const$. An exception is only the range between
$10^{15}\eV$ and $10^{17}\eV$, where horizontal air shower observations are
still dominated by the atmospheric background and underground experiments
affected by earth shadowing.

\sect{APPENDIX:Fermi}{Time scales for Fermi acceleration}

The time scale for first order Fermi acceleration at parallel shock fronts
(defined as having the flow direction parallel to the magnetic field lines, 
with perpendicular and oblique shocks defined accordingly)
is given by \cite{Dru83}
\eqn{tacc:1st}  
\tacc = \frac{\chi_\beta}{\chi_\beta-1}\;\frac{\rgyr}{c \bsh^2}
\Big(y_-+\chi_\beta y_+\Big) 
\quad,
\text
where $\bsh$ is the velocity of the shock in the comoving frame of the
unshocked fluid, and $y_-$ and $y_+$ are the ratios of diffusion coefficients
parallel to the magnetic field, $\kpar$, to the Bohm diffusion coefficient,
\viz, $y = 3 \kpar/\rgyr c \le 1$, taken in the regions upstream and
downstream from the shock, respectively. The velocity jump at the shock in
its comoving frame, $\chi_\beta = \beta_-/\beta_+$, satisfies $\chi_\beta\le
4$ in nonrelativistic shocks (for an ideal gas with specific heat index of
$\frac53$ \cite{Dru83}), and is $\chi_\beta= 3$ in the ultra-relativistic
limit \cite{KS87}. Assuming $y_+\sim y_-\equiv y$, we find $\thF \sim
\bsh^{-2} y$ since $\bsh\equiv\beta_-$ by definition. For parallel shocks,
there is no upper limit on the value of $y$, which can be interpreted as a
measure for the strength of the turbulence in the magnetic field, $y\approx
(B/\delta B)^2$. In perpendicular shocks, one can show that $\thF\sim
\bsh^{-2} y/(1+ y^2)$, and $y$ is limited from kinetic theory and isotropy
requirements by $y < \bsh^{-1}$ \cite{Jok87}.  In ultra-relativistic shocks,
$\bsh \approx 1$, \eq{tacc:1st} is not strictly valid, but numerical
simulations for both parallel and oblique relativistic shocks suggest
$\thF\sim1$ \cite{EJR90/BO96}, in correspondence with the result of
\eq{tacc:1st} for $\bsh=y=1$. The time scale for second order Fermi
acceleration is given by \cite{Kul79}
\eqn{tacc:2nd}
\tacc \approx \frac{3 c \rgyr}{\vA^2}\left(\frac{\delta B}{B}\right)^{-2}\quad.
\text
where $\vA$ is the Alv\'en speed in the plasma.  Introducing $y = (B/\delta
B)^2$ as above, this yields $\tacc \sim y \bA^{-2} \tgyr$, with $\bA =
\vA/c$. The condition for efficient scattering of the particles is $\bA^2\ll
y$ \cite{MelroseII}, which directly translates to $\thF \sim y \bA^{-2} \gg 1$
for second order Fermi acceleration.

Hence, $\tacc\sim \tgyr = 2\pi \rgyr/c$, or $\thF\sim 1$ gives a reasonable
lower limit for the acceleration time scale of any kind of Fermi
acceleration. It may be reached for acceleration in relativistic shock waves;
in most cases, however, factors $\thF\gsim 10$ would be more realistic.
Throughout the paper, $\thF$ is treated as constant, \ie, which assumes that
the diffusion coefficient is proportional to the Bohm diffusion
coefficient. This is not true in other turbulence spectra, \eg, Kolmogorov
turbulence, where $\thF\sim 1$ at the maximum energy implies $\thF\gg 1 $ at
lower energies. While this can be important for the comparison of
electron and proton acceleration time scales \cite{BS87}, it does not affect
too much our results near the maximum proton energy.

\sect{APPENDIX:AGNcore}{VHE/UHE neutrino emission from AGN cores}

The first models which predicted considerable VHE/UHE neutrino fluxes from
AGN assumed particle acceleration at shocks in the accretion flow onto the
putative central black hole \cite{SDSS91,SB92,PS92}. By assuming in these
models that the power-law $2{-}10\keV$ X-ray emission observed from AGN is
produced by $\piO$-decay from $p\gamma$ and $pp$ interactions, and assuming
that the observed diffuse X-ray background is entirely due to AGN, one can
estimate the corresponding neutrino flux arising from $pp$ and $p\gamma$
interactions.  While the assumption of shocks in the accretion flow near the
black hole is more speculative than in jets (unlike the shocks in extended
jets, the inner portions of AGN have never been imaged with sufficient
angular resolution to infer shocks), core shocks are still the most cited
class of models used to estimate expected event rates in neutrino
experiments.  Thus, it useful to investigate such models for self-consistency
in the face of the pion and muon cooling effects discussed in this paper,
which were not considered in the published results (see also more recent
papers, \eg, \cite{SP94,SS96}).

In principle, we could incorporate these models into our general discussion,
because the relevant sizes and time scales in AGN cores are also limited by
variability. However, we will not write down here all the observational
quantities, but use rather the physical parameters applied in the original
papers. The only relevant quantity in the expression of the critical energy
for pion and muon cooling (\Eqs{theory:Enumax:pimu:syn}) is the value of the
magnetic field, since AGN core models are obviously not Doppler boosted
($\cD=1$). Assuming equipartition of the magnetic field and radiation energy
densities, the magnetic field is generally taken as $B\sim [10^3\G]
\cL_{45}^{-1/2}$, where $\cL_{45}$ is the UV luminosity of the AGN.  (Note
that the inverse dependence of the equipartition field on the luminosity
arises from the fact that the luminosity scales with the distance of the
shock from the Black Hole, and thus with the linear size of the acceleration
region, as $\cL\propto R$, leading to $\uph\propto \cL^{-1}$.)  Applying
\Eqs{theory:Enumax:pimu:syn} this leads to $\Enumaxobsmusyn \sim [10^{16}\eV]
\cL_{45}^{1/2}$, and $\Enumaxobspisyn \sim [10^{17}\eV] \cL_{45}^{1/2}$.
Disregarding pion and muon cooling, the models of Stecker et
al. \cite{SDSS91,SS96} predict a flat single-AGN neutrino power spectrum up
to $[2\mal10^{17}\eV]\cL_{45}^{1/2}$ followed by an exponential cutoff;
Protheroe and Szabo \cite{PS92,SP94} find essentially the same result. The
model of Sikora and Begelman \cite{SB92} predicts a sharp cutoff at about
$10^{15}\eV$ due to their more conservative assumption for the acceleration
time scale, $\thF \sim 100 \bsh^{-2}$, rather than $\thF \sim \bsh^{-2}$ as
assumed in the other models. The latter is the only model which is not
modified by considering pion and muon cooling, while for the other models we
see that the electron neutrino spectrum and $50\%$ of the muon neutrino
spectrum, arising from muon decay, steepen more than one order of magnitude
lower in energy that previously assumed.

For the prediction of detector event rates, the integrated diffuse neutrino
background contributed by all AGN needs to be determined. Taking simple step
functions as approximations for both the steepening induced by muon cooling
and the exponential cutoff induced by the maximum proton energy, and noting
that the dependence of the maximum energy on the AGN X-ray luminosity remains
the same, we can derive a relation which allows to transform the result
obtained disregarding muon cooling into the result expected when this effect
is considered. The ratio of the integrated neutrino number flux, $\cF(\cE)$,
to the unmodified single source spectrum, $f_0(\cE)$, for $\cE$ between the
cutoff energies of the least and most luminous AGN, can thus be written as
\eqn{AGNcore:trans}
\cQ(E) \equiv \frac{\cF(\cE)}{f_0(\cE)} = \int_{\cL_{\rm min}}^{\cL_{\rm
	max}} \lumAGN(\cL) \Hvs(\hat\cE\cL_{45}^{1/2} - \cE) \;d\cL \quad,
\text
where $\lumAGN(\cL)$ is the AGN luminosity function, and $\Hvs(x)$ is the
Heaviside step function, $\Hvs(x)=1$ for $x>0$ and $\Hvs(x) = 0$
otherwise. Replacing $\hat\cE \sim 2\mal 10^{17}\eV$, as originally assumed,
by $\hat\cE' = 10^{16}\eV$ as obtained from muon cooling, obviously leads to
the relation $\cQ'(\cE) = \cQ(\cE\hat\cE/\hat\cE') \approx 0.05\cQ(\cE)$,
since $\cQ \nearprop \cE^{-1}$ \cite{SDSS91,PS92}. Since $\cQ'/\cQ =
\cF'/\cF$, the diffuse flux of electron neutrinos (as well as muon neutrinos
from muon decay) is reduced to about $5\%$ of the value previously obtained,
independent of the luminosity function used. Additionally, the exponential
cutoff of the diffuse background, corresponding to the cutoff of the most
luminous quasars, sets in already below $10^{17}\eV$ rather than at
$10^{18}\eV$. Similarly, the diffuse muon neutrino flux from pion decay is
reduced to about $50\%$ of the original value, so that the total VHE muon
neutrino flux drops by about a factor of $3$ and cuts off at $10^{18}\eV$.

Clearly, a more detailed calculation is required to obtain reliable flux
rates under consideration of detailed spectral modification induced by pion
and muon cooling, but our approximate results already show that the effect is
important. In particular, we expect no considerable contribution from AGN
cores to the electron neutrino spectrum in the energy range interesting for
horizontal air shower measurements. No change of the predicted fluxes is
expected in the energy range relevant for deep underwater or ice Cherenkov
detectors, like Lake Baikal or AMANDA. In the interesting intermediate range,
in particular relevant for the event prediction for proposed radio Cherenkov
detectors, we obtain a moderately lower flux of muon neutrinos, and a
severely reduced contribution of electron neutrinos. Since the model
prediction are upper limits (constrained by the diffuse X-ray background),
the drop in the rates cannot be balanced by adjusting astrophysical
parameters.

\end{appendix}

\ifrevtex\bibliographystyle{prsty}
\else\bibliographystyle{unsrt}\fi
\bibliography{agn,grb,neutrino,uhecr,physics}

\ifsubmit 

\sect{APPENDIX:notation}{Symbol table}
\newpage 

\tighten
\settab{tb}{symbols}
\tabcap{Index of frequently used symbols.}
\begin{tabular}{lp{0.46\textwidth}@{\quad}p{0.24\textwidth}l}
symbol 		& meaning	& definition/relations 	& introduced in\\\hline
$\mx$, $\rx$, $\gammax$ 
	& particle mass, classical radius, and Lorentz factor \rule{0pt}{2ex} 
	[$\SS\bullet{=}p,\pi,\mu$] 
	& $\rx = e^2/\mx c^2$
	& general\\ 
$\taux$, $\tauxRF$  
	& unstable particle lifetime [$\SS\bullet{=}n,\pi,\mu$] 
	& $\taux = \gamma_\bullet \tauxRF$ & \RefSec{phothad:pcool} \\  
$\gsprod$, $\gsdec$ 
	& Lorentz factor of secondary particles at production,
	decay\tablenotemark[2] 
	& $\gmudec{\le}\gmuprod{\approx}\gpidec{\le}\gpiprod{\approx}\gamma_p$
	& \RefSec{phothad:pcool} \\
$\Ngtot$, $\Ngb$, $\epsb$, $a$ 
	& total photon density, density and power law index above $\epsb$
	& $d\Ngtot = \eps^{-a}\,d\eps$, $\eps>\epsb$  
	& \RefSec{phothad:pcool}\\
$\eth$ 
	& threshold photon energy in proton RF for $\pi$-production  
	& \Eq{APPENDIX:phothad:Etapi}
	& \RefSec{phothad:pcool}\\
$\gb$, $\gbt$ 
	& characteristic Lorentz factor for power law approximation 
	& $\gb = \eth/2\epsb$, $\gbt = \gb/\cD$
	& \RefSec{phothad:pcool}\\
$\tpib$, $\Etapi$
	& pion production cooling time for $\gamma_p=\gb$, effective
	inelasticity weighted cross section for power law photon spectrum
	& $\tpib = [c\Etapi\Ngb]^{-1}$,\newline
	\Eqs{phothad:tpi}, \RefEq{theory:tpib:lum}\ 
	\RefEq{APPENDIX:phothad:Etapi} 
	& \RefSec{phothad:pcool}\\ 
$B$, $\omBx$ 
	& magnetic field, particle cyclotron frequency 
	& $\omBx = eB/\mx c$ 
	& \RefSec{phothad:pcool}\\
$\tsynx$, $\tad$ 
	& particle synchrotron [$\SS\bullet{=}\pi,\mu,p$] and adiabatic cooling
	time\tablenotemark[1] 
	& \Eq{phothad:tsyn}, $\tad = 2|B/\dot B|$ 
	& \RefSec{phothad:pcool}\\ 
$\tpg$, $\tpi$, $\tescn$, $\tBHp$ 
	& total and specific photohadronic cooling times\tablenotemark[1]
	& $\tpg^{-1} = \tpi^{-1}+\tescn^{-1}+\tBHp^{-1}$
	& \RefSec{phothad:pcool} \\
$\tpcool$ 
	& total proton cooling time scale 
	& $\tpcool^{-1} = \tad^{-1}+\tsynp^{-1}+\tpg^{-1}$ 
	& \RefSec{phothad:pcool} \\
$\fpg$, $\fsyn$, $\fad$, $\fmax$ 
	& rate of proton cooling relative to pion production 
	& \Eq{phothad:tcool/tpi} 
	& \RefSec{phothad:pcool} \\
$\Xnu$, $\Xnus$    
	& efficiency for neutrino production, from specific
	decay\tablenotemark[2] 
	& \Eq{phothad:Xnu} 
	& \RefSec{phothad:pcool} \\
$\Upint$, $\bp$, 
	& total {\em injected} proton energy density, bolometric correction
	factor 
	& \Eq{phothad:Up}
	& \RefSec{phothad:specshape}\\
$\gpmax$, $s$ 
	& maximum proton Lorentz factor, power law index
	& $d\dot N_p\propto\gamma_p^{-s}\,d\gamma_p$, 
	$\gamma_p\lsim\gpmax$
	& \RefSec{phothad:specshape} \\
$\bar L_\nu(E_\nu)$, $\Enumax$, $q$
	& neutrino emission spectrum, cutoff energy, local spectral index 
	& \Eqs{phothad:specnu}
	& \RefSec{phothad:specshape} \\
$\Gamma$, $\bflow$, $\phiview$, $\cD$ 
	& bulk Lorentz factor, velocity in units of $c$, viewing angle and
	Doppler factor of the emission region in observers frame 
	& $\bflow = \sqrt{1-\Gamma^{-2}}$,\newline 
		$\cD = [\Gamma(1-\bflow\cos\phiview)]^{-1}$ 
	& \RefSec{theory}\\ 
$\cT$, $\Lb$, $\bL$ 
	& observed duration of transient, luminosity at $\veps=\epsbobs$ 
	& $T = \cT\cD$, $\Lbobs = \Lb\cD^4$, \Eq{theory:tics}
	& \RefSec{theory}  \\ 
$\Tcr$, $\Tinj$, $\Trad$ 
	& transient crossing, proton injection and radiative time scale 
	& \Eq{theory:Ttot}
	& \RefSec{theory:times:causality}\\ 
$R$, $\Rpar$, $\Rperp$ 
	& linear size of transient emitter, ${\SS\|}=$ in line of sight, 
	$\SS\perp=$ projected 
	& $\Rpar = c\Tcr$ 
	& \RefSec{theory} \\
$\rhoT$, $\xL$
	& geometric correction factors
	& \Eq{theory:Rlim}, \RefEq{theory:lum}
	& \RefSec{theory:times:causality}\\
$\rgyr$, $\tgyr$ 
	& proton Larmor radius and time 
	& $\rgyr = E_p/eB$, $\tgyr = 2\pi\rgyr/c$ 
	& \RefSec{theory:Epmax}\\
$\tacc$, $\thF$ 
	& acceleration time scale, normalized to Larmor time 
	& $\tacc = \thF \tgyr$ 
	& \RefSec{theory:Epmax}\\
$\alpha$, $\bex$, 
	& magnetic field decay parameter, expansion velocity of transient 
	& $B\propto R^{-\alpha}$, $\bex = \dot R/c$ 
	& \RefSec{theory:Epmax:larmor} \\
$\Enumaxobsgyr$, $\rhoL$ 
	& Larmor limit for neutrino energy, correction factor 
	& \Eq{theory:Enumax:Larmor},
	$\rhoL{\lsim}\min(\frac13,\frac1{\pi\thF\alpha\bex})$  
	& \RefSec{theory:Epmax:larmor}\\
$\Enumaxobspsyn$, $\Enumaxobspg$
	& neutrino cutoff energy limited by proton cooling\tablenotemark[1]
	& \Eqs{theory:Enumax:psyn}, \RefEq{theory:Enumax:pg} 	
	& \RefSec{theory:Epmax}\\
$\omdecs$ 
	& characteristic frequency for secondary particle decay\tablenotemark[2]
	& $\omdecs = \frac32 \sqrt{c/\tausRF\rs}$ 
	& \RefSec{theory:pimu} \\
$\Enumaxobssad$, $\Enumaxobsssyn$
	& critical neutrino energies for secondary particle
	cooling\tablenotemark[1]\tablenotemark[2] 
	& \Eqs{theory:Enumax:pimu:ad}, \RefEq{theory:Enumax:pimu:syn}
	& \RefSec{theory:pimu} \\
$\Enudecs$
	& neutrino energy in decay frame\tablenotemark[2]
	& $\Enus = \gs\Enudecs$ 
	& \RefSec{theory:pimu} \\
$\UpsT$, $\XiT$ 
	& dimensionless characteristic parameters of transient 
	& \Eqs{theory:UpsT}, \RefEq{theory:XiT}
	& \RefSec{theory}\\
$\uB$, $\ue$, $\uph$ 
	& magnetic, electron and photon energy density in emission region
	& $\uB = B^2/8\pi$, $\ue\sim\uph$
	& \RefSec{theory:modind:parspace} \\
$\eqBg$, $\eqpB$
	& energy equipartition parameters
	& \Eq{theory:eqBg}, $\eqpB=\Upint/\uB$
	& \RefSec{theory:modind:parspace} \\
$\Dstar$, $\Bstar$, $\gpstar$
	& ``star-point'' parameters\tablenotemark[3], all specific proton
	cooling times ${=}\tacc$  
	& \Eqs{theory:starpoint}
	& \RefSec{theory:modind:parspace} \\
$\Enuobsxlim(\cD)$, $\Bxsyn$
	& neutrino energy limit and corresponding $B$ for given $\cD$
	[$\SS\bullet{=}p,\decsym$]\tablenotemark[2]
	& \Eqs{theory:Bpsyn}--\RefEq{theory:Enumax:D:pimu}
	& \RefSec{theory:modind:uplim} \\
$\LRnulimx$
	& relative neutrino luminosity for $\Enumaxobs=\Enuobsxlim$
	& \Eqs{theory:LRnumax:psyn}, \RefEq{theory:LRnumax:ssyn}
	& \RefSec{theory:modind:efficiencies}\\
$\Etottrans$, $\eqBgmax$
	& total energy budget of transient, corresponding maximum $\eqBg$
	& \Eq{theory:eqBgmax}
	& \RefSec{theory:modind:efficiencies} \\
\end{tabular}
\tablenotetext[1]{Cooling processes are: adiabatic losses ($\rm\SS ad$),
synchrotron radiation ($\rm\SS syn$), photohadronic interactions ($\SS
p\gamma$), (charged) pion production ($\SS \pi,\pi^\pm$), Bethe-Heitler
$e^\pm$ pair production ($\rm\SS BH$), neutron escape ($\rm\SS esc,n$).} 
\tablenotetext[2]{The subscript ``$\SS\decsym$'' denotes secondary particles in
hadronic cascade here and throughout the paper  [$\SS\decsym=\pi^\pm,\mu$]}
\tablenotetext[3]{The superscript ``*'' generally denotes quantities taken at
the star-point of the parameter space, \ie, $\cD=\Dstar$, $B=\Bstar$, and
$\gamma_p = \gpstar$.}
\text

\sect{astro}{Figures}
\newpage

\setfig{p}{AGN:parspace}
\epsfxsize=\textwidth\epsfbox[17 365 593 650]{blazar.ps} 
\figcap{\capsize Dominant cooling processes and neutrino spectral shapes for
AGN jets --- {\bf Left:} Parameter space, with the ``star-point'', denoting
equal cooling time scales at the maximum energy, indicated. The shaded
regions correspond to the dominant cooling process at the maximum proton
energy: (A) Larmor limit or adiabatic cooling; (B) photohadronic cooling; (C)
and (D) synchrotron cooling, where (D) marks the region where photohadronic
cooling dominates for a part of the energy spectrum. Also shown are the
positions of three observed AGN flares: $\exval{1}$ Mkn 421, April 26, 1995 
($\cT_4=10$, $\LXAGN=0.5$) \cite{BAB+96}; $\exval{2}$ Mkn 421, May 7, 1996
($\cT_4=0.1$, $\LXAGN=0.9$) \cite{Schub97}; $\exval{3}$ Mkn 501, April 16,
1997 ($\cT_4=3$, $\LXAGN=2.0$) \cite{PVT+98}. Central positions assume $\uB =
\uph$, black triangles correspond to $\uB=100 \uph$, diagonal errors indicate
the range of possible Doppler factors (see text). Numbers in diamonds
assiciate data points to the corresponding delimiting lines of muon cooling
(black) and pion cooling (white); secondary particle cooling is relevant in
the parameter space region above these lines. {\bf Right:} Schematic
representation of the shapes of neutrino spectra (time integrated power per
logarithmic interval of energy), $\ln \LTnuobs(\Enuobs)$ vs. $\ln \Enuobs$,
corresponding to regions (A)--(D). Break energies due to changes of the
dominant proton cooling process are indicated (\cf\ \RefSec{phothad:pcool}),
possible additional breaks due to secondary particle cooling are omitted for
simplicity (\cf\ \RefSec{theory:pimu:specmod}).}
\text

\newpage

\setfig{p}{GRB:parspace} \centerline{\epsfxsize=0.9\textwidth\epsfbox[0 100
555 780]{grb.ps}} \figcap{\capsize Dominant cooling processes and spectral
shapes for neutrino production in Gamma-Ray Bursts --- {\bf Upper left and
lower panel:} Parameter space for three different parameter sets, all
assuming $\thF=\bGRB=1$, and $\alpha=2$: (A) canonical case, $\LGRBobs =
\TGRBobs = 1$; (B) short intrinsic flares, $\cT=10\ms$, and $\LGRBobs=1$; (C)
extreme case for bright afterglow burst, $\cL=3\mal 10^{52}\erg\scnd^{-1}$
and $\cT=30\scnd$ (which implies an isotropic bolometric photon energy
$\bL\Lbobs\cT = 10^{55}\erg$, requiring $\Omfrac\ll1$). The central line
divides regions where photohadronic cooling dominates adiabatic cooling (left
from the line), and vice versa (right from the line), the star-point of equal
proton cooling times at the maximum energy is indicated. Shaded regions
correspond to spectral shapes produced by the subsequent change of dominant
proton and secondary particle cooling processes: (1) adiabatic/photohadronic
cooling dominant up to $\Enumaxobs$; (2) $\Enumaxobssad<\Enumaxobs$; (3)
$\Enumaxobssad < \Enumaxobsssyn < \Enumaxobs$; (4)
$\Enumaxobsssyn<\Enumaxobs$; (5) $\Enumaxobsssyn<\Enubobspsyn<\Enumaxobs$. In
each figure, the upper part corresponds to neutrinos from pion decay and the
lower part for neutrinos from muon decay. Also indicated are the regions
allowing UHE cosmic ray production up to $\EpUHE$ (black chain lines), and
the region corresponding to the neutrino fluxes predicted in
Ref.~\cite{WB97}, for $\eqpB=1$ (white hatched line and arrows). {\bf Upper
right}: Schematic representation of corresponding spectral shapes, $\log
\LTnuobs(\Enuobs)$ vs. $\log \Enuobs$, corresponding to regions (1)--(5). The
lower break indicates the spectral change at about $100\TeV$ due to the
change of the photon target spectrum at $\gamma_p\sim \gb$ (see
Ref. \cite{WB97}), which is not discussed in this paper.}
\text

\fi 
\end{document}